\documentclass[12pt]{article}

\usepackage{graphics,psfrag}
\usepackage{amsthm,amssymb,epsfig,amsmath,euscript,array,cite}
\usepackage{color}
\usepackage{graphicx}

\newcommand{\be}{\begin{equation}}
\newcommand{\ee}{\end{equation}}
\def\bea{\begin{eqnarray}}
\def\eea{\end{eqnarray}}
\newcommand{\eq}[1]{(\ref{#1})}
\def\nn{\nonumber}

\newcommand{\beq}{\begin{equation}}
\newcommand{\eeq}{\end{equation}}
\newcommand{\ben}{\begin{eqnarray}}
\newcommand{\een}{\end{eqnarray}}
\newcommand{\bes}{\begin{subequations}}
\newcommand{\ees}{\end{subequations}}
\newcommand{\blg}{\begin{align}}
\newcommand{\elg}{\end{align}}

\newcommand{\cN}{{\cal N}}
\newcommand{\cC}{{\cal C}}

\newcommand{\cL}{{\cal L}}

\newcommand{\vev}[1]{{\left< {#1} \right>}}

\newcommand{\red}[1]{\textcolor{red}{#1}}

\newcommand{\blue}[1]{\textcolor{blue}{#1}}

\newcommand{\startappendix}{
\setcounter{section}{0}
\renewcommand{\thesection}{\Alph{section}}}

\newcommand{\Appendix}[1]{
\refstepcounter{section}
\begin{flushleft}
{\large\bf Appendix \thesection: #1}
\end{flushleft}}

\def\N{{\cal N}}

\def\one{\mbox{1 \kern-.59em {\rm l}}}

\def\a{\alpha}

  \def\D{\Delta}  
\def\e{\epsilon}

\def\l{\lambda}

 \def\P{\Pi}

\def\s{\sigma}  \def\S{\Sigma}
\def\t{\tau}
\def\th{\theta}

 \def\cB{{\cal B}} \def\cC{{\cal C}}
  \def\cF{{\cal F}}
 \def\cH{{\cal H}} 
  \def\cL{{\cal L}}
 \def\cN{{\cal N}} \def\cO{{\cal O}}
  
 \def\cT{{\cal T}}

\begin{document}

\hfill{WITS-CTP-089}

\vspace{14pt}

\begin{center}

{\Large \bf
Probing strongly coupled anisotropic plasma
}
\vspace{18pt}

{\bf
 Dimitrios Giataganas
}

{\em
National Institute for Theoretical Physics,\\
School of Physics and Centre for Theoretical Physics,\\
University of the Witwatersrand,\\
Wits, 2050,\\
South Africa
}

{\small \sffamily
dimitrios.giataganas@wits.ac.za
}

\vspace{18pt}
{\bf Abstract}\end{center}

We calculate the static potential, the drag force and the jet quenching parameter in strongly coupled anisotropic $\cN=4$ super Yang-Mills plasma. We find that the jet quenching is in general enhanced in presence of anisotropy compared to the isotropic case and that its value depends strongly on the direction of the moving quark and the direction along which the momentum broadening occurs. The jet quenching is strongly enhanced for a quark moving along the anisotropic direction and momentum broadening happens along the transverse one. The parameter gets lower for a quark moving along the transverse direction and the momentum broadening considered along the anisotropic one. Finally, a weaker enhancement is observed when the quark moves in the transverse plane and the broadening occurs on the same plane.
The drag force for quark motion parallel to the anisotropy is always enhanced. For motion in the transverse space the drag force is enhanced compared to the isotropic case only for quarks having velocity above a critical value. Below this critical value the force is decreased. Moreover, the drag force along the anisotropic direction is always stronger than the force in the transverse space. The diffusion time follows exactly the inverse relations of the drag forces. \\
The static potential is decreased and stronger decrease observed for quark-antiquark pair aligned along the anisotropic direction than the transverse one. We finally comment  on our results and elaborate on their similarities and differences with the weakly coupled plasmas.

\setcounter{page}0
\newpage

\section{Introduction}

Last years there is a lot of effort to understand the heavy-ion collisions and the quark-gluon plasma (QGP). The Relativistic Heavy Ion Collider (RHIC) findings and the following analysis suggests that the QGP is a strongly coupled fluid \cite{scfluid} and therefore the perturbative methods of Quantum Chromodynamics (QCD) are in general not appropriate for describing it. On the other hand there is some progress in Lattice field theory, (see for example \cite{0704.1801}) but further progress is very difficult since in QGP we need to study real-time phenomena. Moreover, a promising approach to study these phenomena is developed with the use of methods within gauge/gravity duality \cite{adscft}, where  an up to date review is in reference \cite{1101.0618}. Using the AdS/CFT it is possible to study several observables and properties of the dual QGP working in the strongly coupled regime. At the moment the studies are mostly in the qualitative level, but still the information that can be extracted is significant.

Although the exact gravity dual of the QCD is not known, and the theories and their dual backgrounds used for QGP calculations usually have different characteristics and properties than QCD (although some of them disappear in high temperatures), there are several important results that appear to have some kind of universality among the different theories. Relativistic hydrodynamics describe well the QGP \cite{0804.4015_0902.3663} and one of the most known results obtained so far is for the prediction of the ratio of shear viscosity over entropy density \cite{Policastro:2001yc}. Moreover several  methods have been developed for calculations of the jet quenching, the drag force and the relaxation time, the static potential and the quarkonia etc. in the dual QGP, which are accompanied with some arguments for  normalizing the results in order to lead to more sensible comparisons with the results obtained from the experiments.

By now there is a large number of papers which study these quantities in the phase where the plasma has become already isotropic and at equilibrium. However the plasma after its creation and for a short period of time is anisotropic both in momentum and coordinate space. For the RHIC energies the elliptic flow of the matter created is described quite well by models that assume that the hydrodynamical behavior is applicable at $\t\lesssim 1 fm$. The conformal viscous hydrodynamics predicts times $\t\sim 2 fm$ but the estimation depends  strongly on the initial conditions (eg. Color Glass Condensate (CGC) or Glauber) and details of plasma hadronization. On the other hand, by considering the collision of two sheets of energy in strongly coupled $\cN=4$ sYM in \cite{yaffe03} has been predicted a thermalization time of order $\sim 0.3 fm$. Therefore, the current estimations of thermalization time vary significantly. 

In this paper we initiate the study of several observables in a dual anisotropic strongly coupled QGP. Such anisotropy that we study here, can be referred to the momentum space and is caused due to locally anisotropic hydrodynamic expansion of the plasma. We point out that the question we answer accurately in this paper is how the observables are modified in the dual $\cN=4$ sYM plasma in presence of anisotropy. Whether or not our results apply to the observed anisotropic QGP, depends mainly on how well the initial isotropic theory, of which our theory here is a consistent deformation, describes the QGP. There are several indications that at least qualitatively the predictions in the isotropic case capture characteristics of the real QGP quite well, so our analysis here as well could capture properties of the anisotropic plasma, and we find that our results are indeed consistent to what is expected. Apart from that, it is very interesting on its own to see how several quantities in context of AdS/CFT are modified in presence of such an anisotropy we consider here.

Recently there is important progress in the anisotropic gauge/gravity dualities. In \cite{janikaniso} it has been found an anisotropic dual geometry with a naked singularity. This geometry was used in \cite{electromagnetic} to study electromagnetic signatures of the dual plasma. In \cite{Erdmengera1} it has been obtained a deviation from the universality of the ratio of the shear viscosity over entropy density when rotational symmetry is spontaneously broken and in \cite{Erdmengera2} several effects due to anisotropy have been studied further in anisotropic superfluids. In \cite{takayanagi} a supergravity solution was found which interpolates between the $AdS_5$ and the Lifshitz-like solution at zero temperature.
In \cite{mateosaniso}  the anisotropic supergravity solutions that were found are regular on and outside the horizon and are generalizations of the zero temperature solution \cite{takayanagi}. The geometry was used in \cite{nsviolation} to find that the longitudinal shear viscosity over entropy bound is violated. The strong coupling isotropization of a large number of anisotropic initial states in the absence
of external sources was also studied recently in \cite{cftiso}. In this paper we continue and extend the studies on the anisotropic gauge/gravity dualities.

Regarding  the QGP anisotropies, it is known that  the existence of them is important for the QGP evolution. For example, a spatial anisotropy which occurs due to the fact that in the heavy ion collision the nuclei have finite area and usually collide off-center causes the hydrodynamic elliptic flow. The flow is associated to the elliptic flow parameter $v_2$ which is defined as the anisotropy of particle production with respect to the reaction plane and is a way to measure how the system responds to the initial spatial anisotropy.
The elliptic flow is generated because the pressure gradient along the impact vector direction on the transverse plane is larger than the one in the transverse direction and the nuclear medium expands preferably along the impact vector direction. Moreover the interaction between the particles, leads to a momentum anisotropy distribution on the reaction plane reflecting the excited medium to the above spatial geometry. The measurement of the elliptic flow provides information for the thermalization times and can be used to constrain the ratio of shear viscosity over entropy density.

Here we are mainly interested for the anisotropy which is created by the rapid expansion of the plasma along the longitudinal beam axis at the earliest times after the collision. The longitudinal pressure to the beam axis is lower than the transverse one and the momenta of the partons along the beam direction are lower than the ones in the transverse space.
By considering boosted hadrons with the same velocity the effective temperature increase with the mass of the hadron species \cite{Adams:2003qm}. These momentum distribution anisotropies cause  plasma color instabilities which are responsible for the isotropization short time and process of the QGP \cite{instabilities}, at least in the weakly coupled regime.

All the different anisotropies mentioned
can occur at the same times. A way to isolate the anisotropy we are interested on is to think the colliding nuclei as having infinite transverse area, or that the collisions are completely central.
After the collisions the partons are produced at the formation time
where the partonic momentum distribution can be supposed to be isotropic. Then a rapid longitudinal expansion of the plasma along the beam line occurs. During this process the longitudinal expansion rate is larger than the parton interaction rate, and the plasma along the longitudinal direction is much colder than the one in the transverse direction. At this stage the pressure along the longitudinal and transverse directions satisfy $P_L<P_T$ and the corresponding momenta $\vev{p_L^2}<\vev{p_T^2}$ in the local rest frame. At the time $\t=\t_{iso}$ the interaction rate becomes equal to the expansion rate the plasma reaches the isotropic phase where the hydrodynamic analysis can be done. This momentum anisotropic plasma have  chromo-Weibel instability, which believed to play important role on the isotropization process at least in the weakly coupled plasmas.

In this paper motivated by the experimental as well as theoretical studies, we initiate the study of several observables in a dual anisotropic plasma. We use a high temperature limit of a static, regular IIB supergravity solution dual to a spatially anisotropic finite temperature $\cN=4$ super Yang-Mills (sYM) plasma \cite{mateosaniso}. The geometry characterized by an anisotropic parameter where its limit to zero is smooth and gives the isotropic undeformed finite temperature $\cN=4$ sYM. This is expected since the supergravity solution can be seen as a deformation of the original solution, where the anisotropy can be though as introduced either by a non-zero number density of dissolved branes that do not extend to boundary and therefore do not add new degrees of freedom, or resulting from a $\th$-term which depends on the anisotropic direction. In the dual gravity the particular $\th$-term corresponds to an axion depending on the anisotropic direction which can be thought as generated by the additional D7 branes.

Using this background we start by calculating the static potential and the static force between a pair of heavy probe quarks Q\={Q}.  We find how the critical length of the pair distance and the values of the static potential depend on the anisotropic parameter along the different directions in the plasma. We compare these results to the isotropic case and then compare with models that study the static potential in the weak coupling regime. Moreover, another motivation for these calculations is that this study might be useful to extract qualitative results for the quarkonia in the anisotropic plasma.

Continuing we calculate the drag force and the diffusion time of a heavy quark moving along different directions of the anisotropic plasma. We derive the analytical results for quarks moving along the transverse and longitudinal directions, compare them each other and to the isotropic case.

Then we study the jet quenching parameter. It's bounds can be measured in the QGP by the radiative energy loss and the parameter itself can be though as a property of the strongly coupled medium. In our 4-dim anisotropic plasma we have three different choices for the transverse momentum broadening. The energetic parton moves along one of the transverse directions and the momentum broadening happens along the anisotropic direction. The second is when the parton moves parallel to the anisotropic direction and the momentum broadening considered in the transverse direction. For the last one the parton moves along the transverse to the anisotropy directions and the momentum broadening is calculated along the other transverse direction. In this case although only the transverse directions are considered, the dependence of the radial metric element on the anisotropic parameter, modifies sightly the result compared to the one in the undeformed theory.  In general we find enhancement of the jet quenching in presence of anisotropy. Our results again compared to the results obtained from other models. Finally, we also discuss the difficulties that arise in our model when we try to give a more precise quantitative prediction using different comparison schemes. These difficulties appear because we are working on a small anisotropy over temperature limit.

In this paper we have tried to present some of the analytic calculations clearly in the Appendices in order to improve the readability of the main text.
The structure of the paper is as follows. In the second section we present the background and the theory we use and how its parameters are related to the parameters of models using anisotropic momentum distribution functions. 
In the following section we investigate the static potential and the static force in the anisotropic dual plasma. This section is supported by the Appendix A, where the generic gravity dual orthogonal Wilson loop calculations are presented. In section 4, the drag force and the quark relaxation time is studied. The appendix B supports this section where the analytical calculations for the drag force are presented for any general background. In section 5, we study the jet quenching. In this section we present in the main text the jet quenching calculation and the approximations done to derive the result for a complete generic background, since the calculation is very interesting and certain approximations done in the derivation need to be tested in our background. Then we derive the results in the anisotropic background and comment on them. In the next section we make an attempt to provide more quantitative predictions of our results using different comparison schemes. We finalize with the discussion section where we also collect our results.

\vspace{-0.8cm}
\section{The model}
\vspace{-0.51cm}
\subsection{The dual geometry}
\vspace{-0.3cm}
The anisotropic background we use is a deformed version of the $\cN=4$ finite temperature sYM \cite{mateosaniso}. The deformation parameter in the field theory is introduced by a $\th$-parameter term depending on the anisotropic direction. It turns out that $\th =2 \pi n_{D7} x_3$, where $x_3$ is the anisotropic gauge theory space coordinate and $n_{D7}$ can be thought as the density of $D7$-branes homogeneously distributed along the anisotropic direction. The $\th$ angle is related to the axion of the type IIB supergravity through the complexified coupling constant of the $\cN=4$ sYM. Therefore in the gravity dual background the anisotropic deformation can be seen as inserted due to existence of axion term depending on the anisotropic direction, or as due to the backreaction of the $D7$-branes. These branes do not add new degrees of freedom to the theory since they do not touch the boundary. They are wrapped on the internal space and the transverse directions to the anisotropy, therefore creating the anisotropy on the deformed $AdS$ geometry.

In the string frame the background is given by
\bea
&&\hskip -.35cm
ds^2 =
 \frac{1}{u^2}\left( -\cF \cB\, dx_0^2+dx_1^2+dx_2^2+\cH dx_3^2 +\frac{ du^2}{\cF}\right)+ {\cal Z} \, d\Omega^2_{S^5}\,.
 \label{metric} \\
&& \hskip -.35cm \chi = a x_3, \qquad \phi=\phi(u) \,,\qquad 
\label{chi}
\eea
where $a$ is the anisotropic parameter with units of inverse length, $\phi$ is the dilaton, $\chi$ is the axion depending linearly on the $x_3$ coordinate. 
The anisotropic direction is considered to be the $x_3$ and the functions $\cF, \cB, \cH$ depend on the radial coordinate $u$ and the parameter $a$. 
The background has also a RR five form but it is not important for our purposes. The analytical form of the functions can be found for small anisotropy compared to the temperature or sufficiently high temperatures, $T\gg a$.
It is enough to consider the expansions of the fields up to second order in $a$ around the black D3-brane solution:
\bea
\cF(u) &=& 1 - \frac{u^4}{u_h^4} + a^2 \cF_2 (u)  +\mathcal{O}(a^4)\\
\cB(u) &=& 1 + a^2 \cB_2 (u) +\mathcal{O}(a^4)\,, \\
\cH(u)&=&e^{-\phi(u)},\quad\mbox{where}\quad \phi(u) =  a^2 \phi_2 (u)  +\mathcal{O}(a^4)\,.
\label{smallae}
\eea
Note that only even powers can appear because of the symmetry $z\to -z$.
By applying asymptotic $AdS$ boundary conditions and requiring $\cF_2$ to vanish at the horizon $u=u_h$, the Einstein equation can be solved giving:
\bea
\cF_2(u)&=& \frac{1}{24 u_h^2}\left[8 u^2( u_h^2-u^2)-10 u^4\log 2 +(3 u_h^4+7u^4)\log\left(1+\frac{u^2}{u_h^2}\right)\right]\,,\cr
B_2(u)&=& -\frac{u_h^2}{24}\left[\frac{10 u^2}{u_h^2+u^2} +\log\left(1+\frac{u^2}{u_h^2}\right)\right] \,,\cr
\phi_2(u) &=& -\frac{u_h^2}{4}\log\left(1+\frac{u^2}{u_h^2}\right)\,.
\eea
We can find the temperature evaluating the following expression at the horizon
\be\label{tuh}
T=-\frac{\partial_u\cF \sqrt{\cB}}{4\pi}\bigg|_{u=u_h}=\frac{1}{\pi u_h}+a^2 u_h \frac{5 \log2-2}{48 \pi} +\mathcal{O}(a^4)
\ee
and solving for the horizon position $u_h$ we get
\be\label{uht}
u_h=\frac{1}{\pi T}+a^2 \frac{5 \log2-2}{48 \pi^3 T^3}+\mathcal{O}(a^4)~.
\ee
As expected the isotropic limit $a\rightarrow 0$ reproduce the well know results of the isotropic black D3-brane solution. The parameters of our background now are the temperature $T$ and the anisotropy $a$.
The energy density per unit volume can be calculated from
\be
s = \frac{A_h}{4 G V_3}~,
\label{entr} \qquad d A_h = \frac{e^{-\frac{\phi_h}{2}}}{u_h^{3}} \, dx\, dy\, dz ~,
\ee
where $d A_h$ is the area element of the hypersurface  $t=\mbox{const}, u=u_h$.
In our case the result reads
\be
s \propto\frac{\pi^2  N_c^2   T^3}{2}+a^2 \frac{ N_c^2  T}{16}+\mathcal{O}(a^4)\,.
\label{entra2}
\ee
The energy and pressures can be found from the expectation value of the stress tensor, where the element  $\vev{T_{00}}$ is the energy, $\vev{T_{11}}=\vev{T_{22}}=P_{x_1 x_2}=:P_\perp$ denote the pressure along the $x_1$ or $x_2$ direction or the transverse plane and $\vev{T_{33}}=P_{x_3}=:P_\parallel$ is the pressure along the anisotropic direction, calling it also longitudinal one.  The analytic expressions read
\bea
E&=& \frac{3\pi^2  N_c^2  T^4}{8}+ a^2
\frac{ N_c^2  T^2}{32}+{O}(a^4)\,,\cr
P_{x_1 x_2}&=& \frac{\pi^2  N_c^2  T^4}{8}+
a^2\frac{ N_c^2  T^2}{32}+{O}(a^4)\,,\cr
P_{x_3}&=& \frac{\pi^2  N_c^2  T^4}{8}-
a^2\frac{ N_c^2  T^2}{32}+{O}(a^4)~. \label{pxyz}
\eea
Therefore for high temperatures the pressure of the plasma along the anisotropic direction is always lower than the one in the other two directions
\be\label{pressaniso}
P_{x_3}<P_{x_1 x_2}~.
\ee

\subsection{Relation of background parameters to anisotropic momentum distribution function}

In weakly coupled anisotropic plasmas a usual technique used for the study of the observables is to consider an anisotropic phase space distribution function. The degrees of freedom in weakly coupled plasmas are split to soft modes that carry momenta of order $g_{YM} T$ and hard ones that carry moment of order $T$. The hard modes are particles that have an anisotropic phase space distribution function.

To make connection of this function with our parameters we can consider a kinematic example where the accelerated beams with nucleons collide along the $x_3$ anisotropic direction. This is the beam-axis direction and along this direction the system expands rapidly initially. The plasma created can be seen as having a distribution function $f(t,\textbf{x},\textbf{p})$ which can be taken homogeneous in position space but anisotropic in momentum space. The anisotropic distribution function can be written as \cite{anisofunction}
\be\label{faniso}
f_{aniso}(\textbf{p})=c_{norm}(\xi) f_{iso}(\sqrt{\textbf{p}^2+\xi (\textbf{p}\cdot \textbf{n})^2})~,
\ee
where the vector $\textbf{n}=(0,0,1)$ is the unit vector along the anisotropic direction and the parameter $\xi$ plays the role of the anisotropic parameter. This distribution represents a stretched or contracted version of the isotropic case since one direction in the momentum space is rescaled.  For $-1<\xi<0$ the distribution is stretched along the anisotropic direction while for $\xi>0$ the distribution is contracted in the anisotropic direction as in Figure 1.
\begin{figure}
\centerline{\includegraphics[height=60mm]{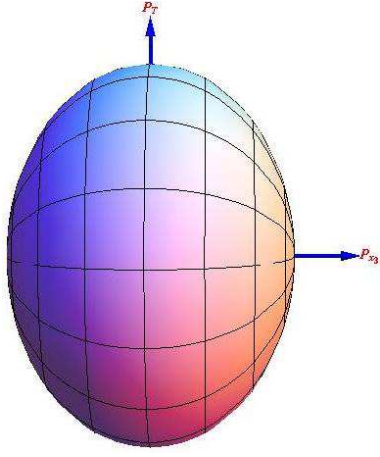}}
\caption{\small{Anisotropy in the momentum space for $\xi>0$.}}
\end{figure}

The parameter $\xi$ is related to the average particle momenta in the transverse $\textbf{p}_T=\textbf{p}-\textbf{n}(\textbf{p}\cdot \textbf{n})$ and longitudinal $p_L=\textbf{p}\cdot \textbf{n}$ to anisotropy directions by
\be\label{xidefine}
\xi=\frac{\vev{p_T^2}}{2 \vev{p_L^2}}-1~,
\ee
where the factor of $2$ appears in the denominator due to the number of the transverse directions. Therefore we see that the anisotropic distribution can be obtained from the isotropic one by removing or adding particles that have large momentum component along the anisotropic direction. 
The anisotropic plasma created after the heavy ion collisions correspond to values $\xi>0$, and this is the region that we work here.

In order to find a qualitative relation between the anisotropic parameter $\xi$ and the parameter $a$ of our supergravity background, we have to express the first one in terms of the pressures of the system. This is done by using the distribution \eq{faniso} and calculating the pressures through the stress energy tensor components. A new parameter $\D$ can be introduced which measures the degree of momentum anisotropy through pressures and defined as
\be
\D:=\frac{P_T}{P_L}-1=\frac{P_{x_1 x_2}}{P_{x_3}}-1~.
\ee
After some algebra $\D$ can be related to $\xi$ \cite{0902.3834} as
\be\label{dxi1}
\D=\frac{1}{2}(\xi-3)+\xi\left(\left(1+\xi\right)\frac{\arctan\sqrt{\xi}}
{\sqrt{\xi}}-1 \right)^{-1}~,
\ee
which in the small $\D$ and consequently small $\xi$ limit, the expression becomes
\be
\lim_{\xi\rightarrow 0}\D=\frac{4}{5}\xi +\cO(\xi^2)
\ee
and in the large $\xi$ limit
\be
\lim_{\xi\rightarrow \infty}\D=\frac{1}{2}\xi +\cO(\sqrt{\xi})~.
\ee
 For our background in the high temperature limit using \eq{pxyz} we get for $\D$
\be\label{da}
\D=\frac{a^2}{2 \pi^2 T^2}~.
\ee
In the range of $T\gg a\Rightarrow\D\ll 1$ we can relate the two anisotropic parameters as
\be\label{xia}
\xi\backsimeq \frac{5 a^2}{8 \pi^2 T^2}~,
\ee
where $\xi$ has to be positive and very small.
The equation \eq{xia} provides the basic connection between the parameters of our supergravity background and the anisotropic momentum distribution functions  \eq{faniso} considered in several field theory models. However it should be noted that this relation obtained only through the pressure anisotropies, since the anisotropic theory we are using here comes from a position $\theta$ dependent angle. Therefore,
the equation \eq{xia}, can be seen as a simple connection of the parameter $a$ and the parameter $\xi$ based  on the pressure anisotropies of the two systems.

Moreover if $f_{iso}$ represents an ideal gas momentum distribution and $\xi$ is small enough, the anisotropic parameter is related to shear viscosity over entropy density and to the 
 proper time of the plasma. For one dimensional Bjorken expansion the analytic relation is \cite{0608270}
\be\label{xieta}
\xi =\frac{10 \eta}{T \t s}~,
\ee
where the anisotropy increases with the expansion rate.
In the following sections we calculate several physical observables, we explain our results and try to qualitatively compare our results with experimental data and the weakly coupled plasma models \footnote{Comparisons using gauge/gravity dualities with the corresponding weakly coupled results in isotropic case have been performed for example in \cite{horo1}.}.

\section{Q\={Q} Static Potential and Static Force in the anisotropic $\cN=4$ plasma}

In this section we study the static potential and the static force in the finite temperature anisotropic dual plasma. In order to do so we are using the analytic equations \eq{staticL} and \eq{staticE} derived in the Appendix A. We will compare the static potential to the isotropic case as well as the potentials along the different anisotropic directions.

We use the usual ansatz for the string world-sheet choosing the static gauge and the $\sigma$ dependence on the radial direction:
\bea
x_0=\t\qquad\mbox{and}\qquad x_p=\sigma,
\qquad u=u(\s)~,\\
\mbox{where}\quad x_p=x_1=:x_\perp\quad\mbox{or}\quad x_p=x_3=:x_\parallel~.
\eea
In the first case we align the Q\={Q} pair along the direction of $x_1$, which is equivalent aligning it along $x_2$. Then the pair is placed along the anisotropic $x_3$ direction. Finally in order to compare with the isotropic finite temperature $\cN=4$ sYM theory we set $a=0$ and calculate the static potential, where the particular analysis is equivalent to the configuration considered in \cite{WLfiniteT}. These are the three different static potentials we study and compare each other.

\begin{figure*}[!ht]
\begin{minipage}[ht]{0.5\textwidth}
\begin{flushleft}
\centerline{\includegraphics[width=70mm]{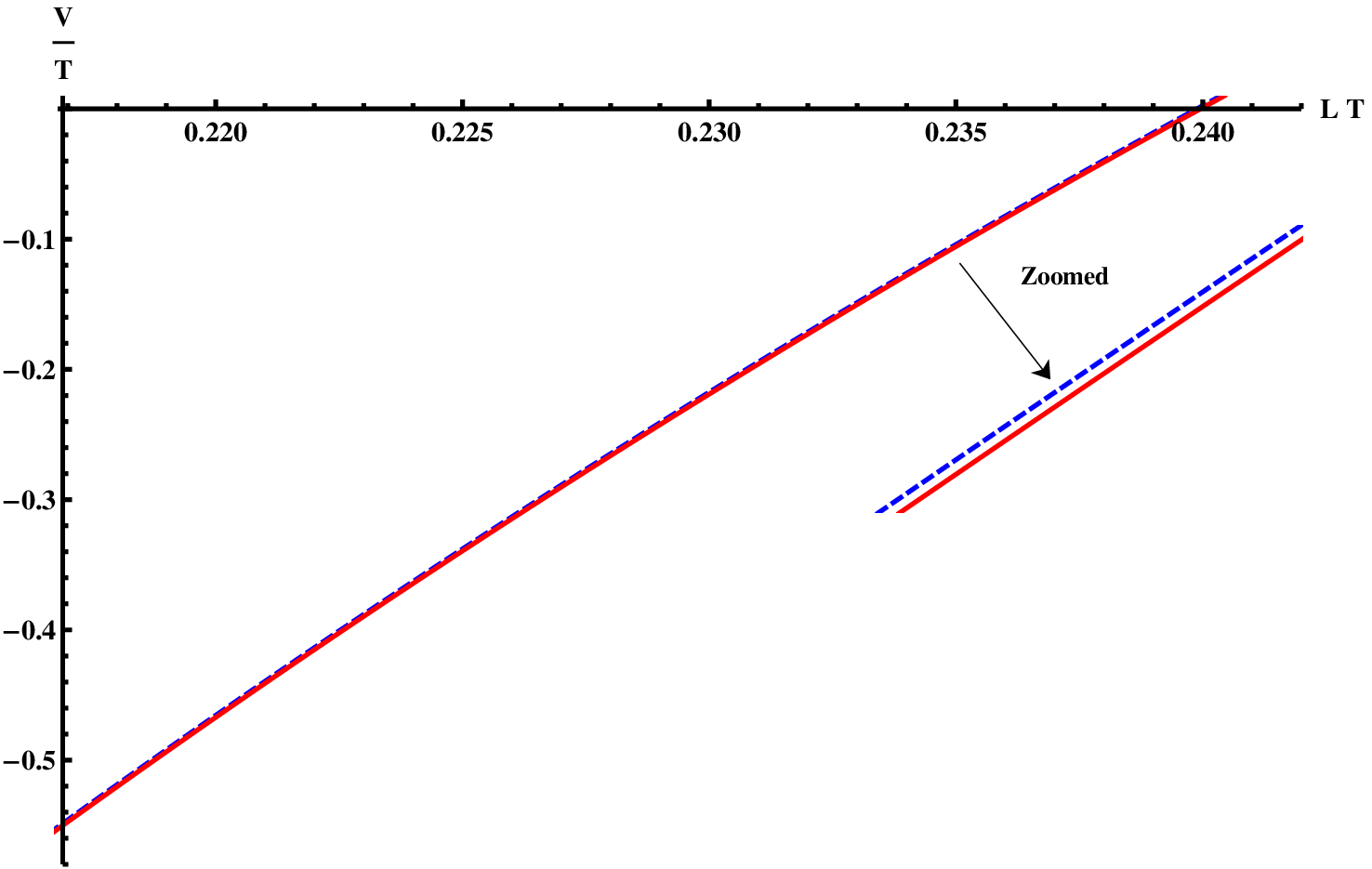}}
\caption{\small{The static potential close to $L_c$ for Q\={Q} pairs aligned along the anisotropic and the transverse direction with respect to $L$. Their relation is $|V_\parallel|<|V_\perp|$. The corresponding critical  lengths are $L_{c\parallel}<L_{c\perp}$. Settings: blue dotdashed line-\blue{$V_\parallel$}, red solid line-\red{$V_\perp$} and $T=3$, $a=0.3 T$.
}}\label{fig:a1}
\end{flushleft}
\end{minipage}
\hspace{0.3cm}
\begin{minipage}[ht]{0.5\textwidth}
\begin{flushleft}
\centerline{\includegraphics[width=70mm]{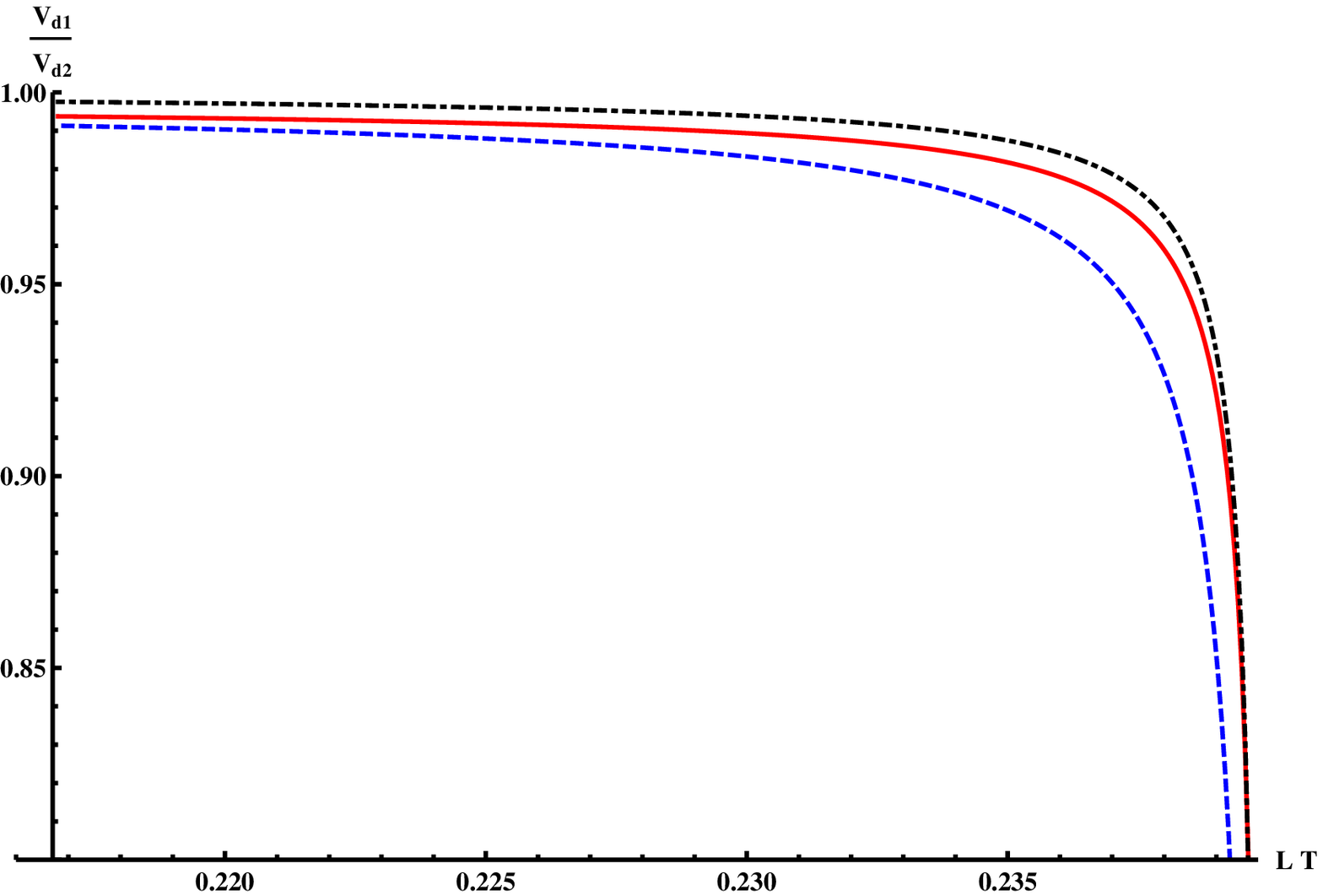}}
\caption{\small{The ratios of the static potentials for pairs aligned along different directions. All the fractions reduce as the pair distance increase and  approach the critical length.
Settings:$V_\parallel/V_\perp$-black dotdashed color, \blue{$V_\parallel/V_{iso}$}-blue dashed line, \red{$V_\perp/V_{iso}$}-solid red line and $T=3$, $a=0.35 T$.}}\label{fig:a2}
\end{flushleft}
\end{minipage}
\hspace{0.3cm}
\begin{minipage}[ht]{0.5\textwidth}
\begin{flushleft}
\centerline{\includegraphics[width=70mm]{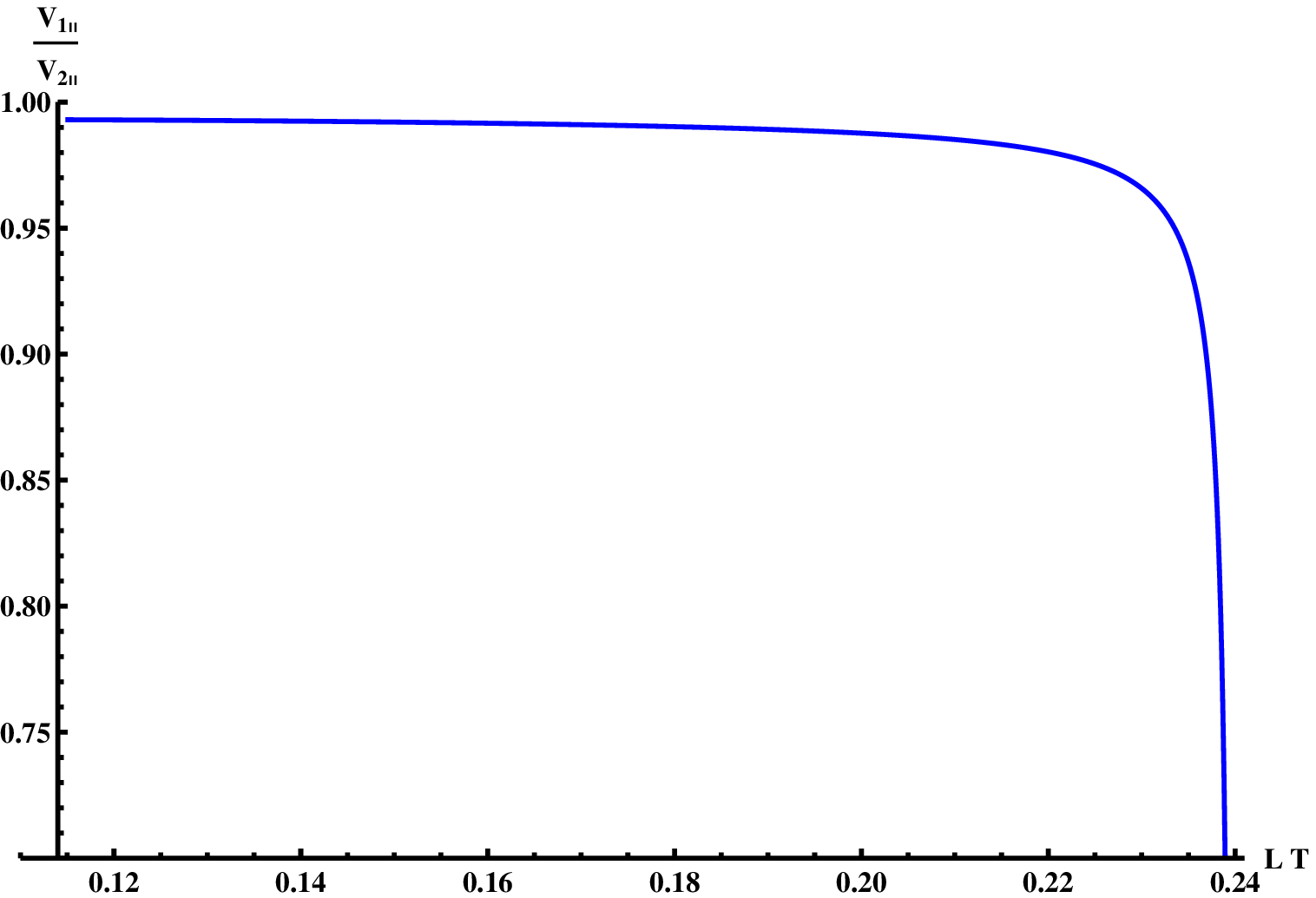}}
\caption{\small{$V_\parallel$ for two values of the anisotropic parameter. The static potential decreases for increasing the anisotropy. Settings: $a_1 = 0.5 T$, $a_2 = 0.01 T$ and $T=3$.\vspace{3.25cm}}}\label{fig:a4}
\end{flushleft}
\end{minipage}\hspace{0.3cm}
\begin{minipage}[ht]{0.5\textwidth}
\begin{flushleft}
\centerline{\includegraphics[width=70mm]{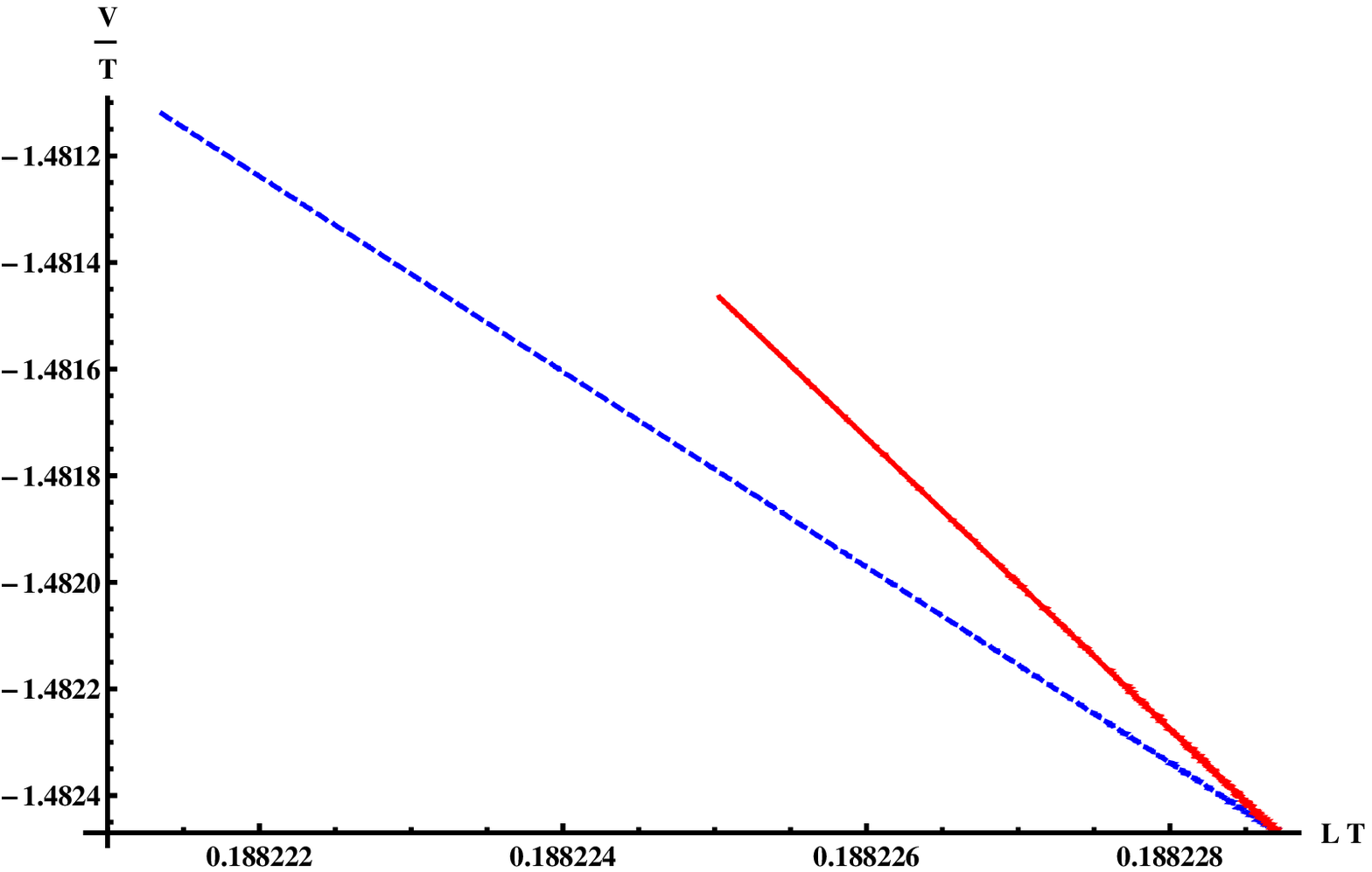}}
\caption{\small{The $V_\parallel$ and $V_\perp$ where the string world-sheet turning point is fixed to $u_0/u_h=0.5$ and $a$ increases. The $a\backsimeq 0$ point is where the lines cross, and as $a$ increases the lines diverge. By comparing the two final points of the curves which correspond to $a\simeq0.5 T$ we see that as $a$ increases the background geometry and the potentials along the parallel direction is affected more than transverse one. Settings: as in Figure \ref{fig:a1}
and $T=3$.}}
\label{fig:figa3}
\end{flushleft}
\end{minipage}
\end{figure*}

Initially we fix the temperature and the anisotropy parameter and consider the static potential of quark pairs with different separation lengths. By applying the formulas derived in the Appendix A we find that the potential for the pairs aligned along the anisotropic direction tends to be slightly weaker compared to transverse alignment as can be seen in Figures \ref{fig:a1} and \ref{fig:a2} {}\footnote{We should comment here that in some of the plots we choose the anisotropic parameter $a$ of order $~0.3 T$ or so. Although smaller values give the same behavior, their effects are not visible in the plots clearly.}:
\be\label{vrelation}
V_{\parallel}<V_{\perp}<V_{iso} \quad\mbox{for}\quad \frac{a}{T}\neq 0 \quad\mbox{and constant}.
\ee
We also find that  the difference between the potentials in two directions gets bigger as the distance of the quarks approach the critical length $L_c$.
Moreover, the critical length is reduced  in  presence of anisotropy as
\be\label{lrelation}
L_{c\parallel}<L_{c\perp}<L_{c~iso} \quad\mbox{for}\quad \frac{a}{T}\neq 0 \quad\mbox{and constant}.
\ee
In order to study the static potential dependence on the anisotropy parameter we keep constant the temperature and vary the anisotropic parameter. Increase of anisotropy leads to decrease of the absolute value of the static potential in any direction compared to the isotropic case (Figure \ref{fig:a4}) and therefore to decrease of the critical length
\be
a\nearrow~\Rightarrow V_{\parallel,\perp}\searrow~\Rightarrow L_{c~\parallel,\perp} \searrow~.
\ee
A more detailed analysis shows that as the anisotropy increases, the deviations of the anisotropic static potentials along the different directions increase (Figure \ref{fig:figa3}).
The anisotropy affects the configuration along the anisotropic direction stronger than the transverse one. This can be seen in Figure  \ref{fig:figa3},
where we fix the temperature, the ratio $u_0/u_h$ and so consequently the length $L$ of the Wilson loop with respect to the horizon position,  and we increase the parameter $a$. Notice that now along the $x$-axis, same lengths $L$ along different directions, correspond to different values of anisotropy
but the first and the last points in the two curves  correspond to the same value of the anisotropy parameter. We see that $V_{\parallel}=V_{\perp}$ for $a\rightarrow 0$, and by increasing $a$ the potentials and the pair distances along the different directions deviate. This happens because the different strength on the modifications of the background geometry along the different directions is reflected to the static potential.

\begin{figure*}
\centerline{\includegraphics[width=77mm]{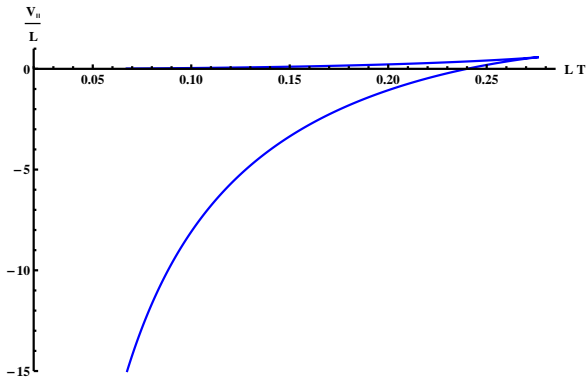}}
\caption{\small{The form of the potential in anisotropic background  is the same as the undeformed $\cN=4$ sYM one. This is not unexpected since the anisotropy introduced by a continuous deformation to $\cN=4$ sYM. Settings: $T=1$ and $a=0.001$.}}\label{fig:a5}
\end{figure*}

Finally, by keeping the anisotropy constant and increase the temperature the static potential in any direction gets lower absolute value and the critical length reduces (Figure \ref{fig:a5}). This is not unexpected since our dual theory is a smooth deformation of $\N=4$ sYM where the same behavior has been observed \cite{WLfiniteT}.

\subsection{Comments and comparison with other models}

To interpret our findings one could think that lower pressure in the parallel direction, which results having less energetic gluons leads to the further screening of the potential. However this seems not to be the case. The enhancement of the anisotropy while keeping constant temperature, results to increase of $P_{x_1 x_2}$ and simultaneous decrease of $P_{x_3}$ as can be seen from \eq{pxyz}, or equivalently leads to contraction of the momentum in the anisotropic direction, by removing for example energetic gluons in that direction. At the same time momentum extension happens along the transverse direction and increase of the relevant pressure. However, here we find that in both directions the absolute value of the static potential is decreased. An other observation that indicates that the static potential is not related directly to the pressures is that at least to $a^2$ the pressures are modified by the same amounts from the initial one. If the force modification between the quarks would depend mainly on the pressure then the magnitude of modification should be almost equal in parallel and transverse directions. This does not happen and it is not expected to happen judging from the differences in the metric in the two directions. The decrease of the static potential is more likely to be related to the increase of energy or entropy density of the system as found in equations \eq{entra2} and \eq{pxyz}.

\begin{figure*}
\centerline{\includegraphics[width=95mm,height=65mm]{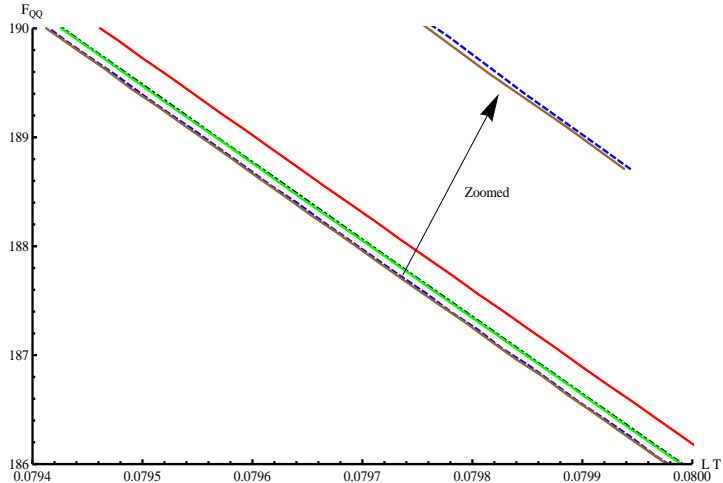}}
\caption{\small{The static force along different directions. The forces in presence of anisotropy are decreased compared to the isotropic plasma with the order \eq{forceqq}. The forces along the transverse and parallel directions for $a=0.35$ are zoomed in, in order to observe their ordering since they are very close. Settings: $T=3$, and the curves starting from bottom; $a=0.35 T$: $F_{Q\bar{Q},\perp}$-brown solid, $F_{Q\bar{Q},\parallel}$-blue dashed; $a=0.30 T$: $F_{Q\bar{Q},\parallel}$-green solid, $F_{Q\bar{Q},\perp}$-black dot-dashed and $F_{iso}$-red solid.  } }\label{fig:statF}
\end{figure*}

Here it should be noted that in general the static potential normally has a constant term which seems to be not physical.
In our analysis we need to take the derivative of the static potential to get rid of the constant term and compare the static forces:
\be
F_{Q\bar{Q}}=\frac{\partial V}{\partial L}~.
\ee
The static force turns out to be decreased in presence of anisotropies, ie. the screening is increased. Further decrease happens as the anisotropy is increasing. However, the decrease here seems to be slightly less in the anisotropic direction that the transverse one, qualitatively following the ordering:
\be\label{forceqq}
F_{Q\bar{Q},\perp}<F_{Q\bar{Q},\parallel}<F_{iso}~.
\ee
These results are plotted in the Figure \ref{fig:statF}.

Notice that at some point in our analysis we have compared the potential in anisotropic background to the isotropic one. When comparing the potentials along the parallel and transverse directions in the anisotropic background for fixed $a/T$ the horizon of the black hole remains at the same position and only the corresponding geometry that the string worldsheets extend is modified. While comparing to the isotropic case the position of the black hole horizon differs but the comparison can be made by identifying the temperatures in the isotropic and anisotropic backgrounds.

Our theory does not have dynamical degrees of freedom in the fundamental representation. Inclusion of quarks in isotropic $\cN=4$ sYM with additional D7 flavor branes beyond the probe limit is expected to lead to screening of the potential at least at scales close to the string breaking one \cite{giataganasSS,qqscreened}. When the anisotropy introduced in a background that contains already dynamical degrees of freedom is not clear from our analysis whether or not the potential will be further reduced. This will depend on the flavors dependence of the anisotropy. If for example the density of flavors turn out to depend on the anisotropy and it is reduced for increasing anisotropy then the static potential could increase since screening will be weaker and the decrease due to gluons found here will be an antagonizing behavior. On the other hand if the density of the flavors will be independent of the anisotropy, then the potential even in presence of flavors will be reduced as we have found here. Additionally, the relation between the static potential along the different directions might get modified after the inclusion of flavors, since their dependence on the anisotropic parameter and therefore their  contribution to screening can not be a priori predicted. Therefore in this paper we have found the behavior of the static potential in presence of anisotropy and matter in adjoint representation, but to predict the results in presence of unquenched dynamical quarks, further analysis is needed.

It is interesting to report in this section some findings of the static potential in the weak coupling regime. In \cite{qqaniso} it has been found  stronger attraction for quark pairs aligned along the direction of the anisotropy than for transverse alignment and that the potential generally gets enhanced in presence of anisotropy. These results are valid on distance scales on the order of the inverse of the Debye mass (while the relevant plots are for $L\sim(0.5,2)m_{D}^{-1}$) and for small anisotropic parameter. In this regime a naive extrapolation shows that already in our configuration we have entered in the deconfined phase. Moreover the authors find that as the separation length of the pair gets reduced the difference between the potential in the transverse and parallel directions is reduced. The static potential results in our strong coupling theory without flavors differ with the ones observed for the weak coupling plasmas. This difference might be generated from the different values of the non-physical constant in the static potential.
However, we point out there exist several differences between the  two models apart of course for the main difference that these findings are in weak coupling.

Before we move on to the drag force analysis we summarize only some of the findings. In this section we have found that the anisotropy in our model results to the decrease of the static potential and affects more significantly the potential for a pair aligned along the anisotropic direction. The static potentials along different directions follow  the relation \eq{vrelation}
and  increase of the anisotropy leads to decrease of the static potential in any direction. The critical lengths $L_{c\parallel,\perp}$ depend on the anisotropy and for increasing anisotropy they reduce \eq{lrelation}. The static force is decreased for increasing anisotropy and along the different directions follows the inequality \eq{forceqq}.

\section{Drag force on the heavy quarks in the anisotropic plasma}

The drag force in an anisotropic plasma depends on the direction of the motion of the heavy external probe quark. We consider a heavy quark moving along the anisotropic direction and then along the transverse direction  with a fixed velocity $v$, in an infinite volume of gluon plasma and fixed finite temperature $T$ \cite{dragf}\footnote{A discussion of the early-time energy loss is in \cite{guijosa1}.}. As usual the force we measure here is the one that needs to be imposed on the quark in order to keep it moving with the constant velocity $v$. This force can be generated for example from a constant electric field, which will balance the backward drag force on the quark generated from the interaction of quark with  gluons and quarks in the plasma.

The velocity $v$ is bounded in order the drag force calculation to be valid. To make sure that the quark indeed loses energy and the generated force from the plasma is a backward drag force, it should move well above the subthermal velocities $v_{subthermal}\sim \sqrt{T/M_Q}$, where $M_Q$ is the mass of the heavy quark. So the low bound in the velocity is $v\gg v_{subthermal}$. Additionally there is an upper bound which should be imposed in order for our calculation to be valid. This is required because for example to keep very large constant speed $v$ through the plasma, a very large electric field is needed which after a critical value would produce q\={q} pairs. There are different ways to calculate the maximum value of the velocity: either by consider the DBI action of a D7-brane that represents the heavy quark and taking a reality condition or by comparing the deceleration forces from drag and vacuum radiation and imposing the first one to be the dominant. The maximum
value  of the velocity in our theory remains approximately similar to the undeformed $\cN=4$ sYM and is:
\be
v^2<1-\left(\frac{\sqrt{\l} T} {M_Q}\right)^4~,
\ee
where  $\sqrt{\l} T/M_Q\ll1$.

In order to describe the trailing string for a motion along the $x_p:=x_{\parallel,\perp}$ directions we use the radial gauge choice:
\be
x_0=\t, \qquad u=\s, \qquad x_p= v \t +f(u)~,
\ee
where in the other directions the world-sheet is localized.
The drag force derivation can be done generically for any background. We derive the force in the Appendix \ref{app:drag}. The final equations we obtain and we use in this section are the equations \eq{dragu0} and \eq{drag}. It is interesting that the drag force depend on the time metric element, on the metric element along the direction where the quark moves and only indirectly to the radial metric element through \eq{dragu0}.

The results obtained in this section are analytical and we have also check them numerically. The radial position $u_0$, where the numerator and denominator change sign simultaneously in \eq{dragxi} in order to keep $f(u)'$ real, can be written as:
\be
u_{0\parallel}=u_{01\parallel}+a^2 u_{02\parallel}~, \qquad u_{0 \perp}=u_{01\perp}+a^2 u_{02\perp}~,
\ee
where $u_{01}$ and $u_{02}$ can in principle depend on all the parameters of the problem, but not on the anisotropic parameter $a$. They are considered for parallel and transverse motion with respect to the anisotropy parameter.
Plugging them in the equation \eq{dragu0} for parallel and transverse motion correspondingly and solving each of them separately and for different orders, we obtain:
\bea
&&u_{01\parallel}=u_{01\perp}=\frac{\left(1-v^2\right)^{1/4}}{\pi  T}=u_{0~iso}~,\\
&&u_{02\parallel}=-\frac{\sqrt{1-v^2}\left(1+\sqrt{1-v^2}\right)+\left(7 v^2-5\right) \log\left(1+\sqrt{1-v^2}\right)}{48 \pi ^3 T^3 \left(1-v^2\right)^{3/4}}~,\\
&&u_{02\perp}=-\frac{ \sqrt{1-v^2}\left(1+\sqrt{1-v^2}\right)+\left(4 v^2-5\right) \log\left(1+\sqrt{1-v^2}\right)}{48 \pi ^3 T^3 \left(1-v^2\right)^{3/4}}~.
\eea
Notice that the $u_{01}$ is equal to the isotropic result as expected and the only difference in $u_{02}$ expressions for different directions is a numerical factor. The drag forces using equation \eq{drag} are given by:
\bea\label{fpar}
\frac{F_{drag,\parallel}}{\sqrt{\l}}&=&-\frac{\pi T^2 v}{2\sqrt{1-v^2}}\\\nn
&-&a^2 \frac{v}{48 \pi}\left( \frac{v^2}{ \left(1-v^2\right) \left(1+\sqrt{1-v^2}\right)} +\frac{2(1-v^2)+ \left(1+v^2\right)  \log\left(1+\sqrt{1-v^2}\right)}{\left(1-v^2\right)^{3/2} }\right)~,\\\label{fper}
\frac{F_{drag,\perp}}{\sqrt{\l}}&=&-\frac{\pi T^2 v}{2\sqrt{1-v^2}}\\\nn
&-&a^2 \frac{v}{48 \pi} \left( \frac{ v^2}{ \left(1-v^2\right) \left(1+\sqrt{1-v^2}\right)}+ \frac{2(1-v^2) - \left(5-4 v^2\right) \log\left(1+\sqrt{1-v^2}\right)}{\left(1-v^2\right)^{3/2}}\right)~.
\eea
The 0th order term in anisotropy is equal to the drag force in the isotropic undeformed background as it should be. The anisotropic correction does not depend on the temperature in any direction and it is consistent with the dimensional analysis.

\begin{figure*}[!ht]
\begin{minipage}[ht]{0.5\textwidth}
\begin{flushleft}
\centerline{\includegraphics[width=70mm]{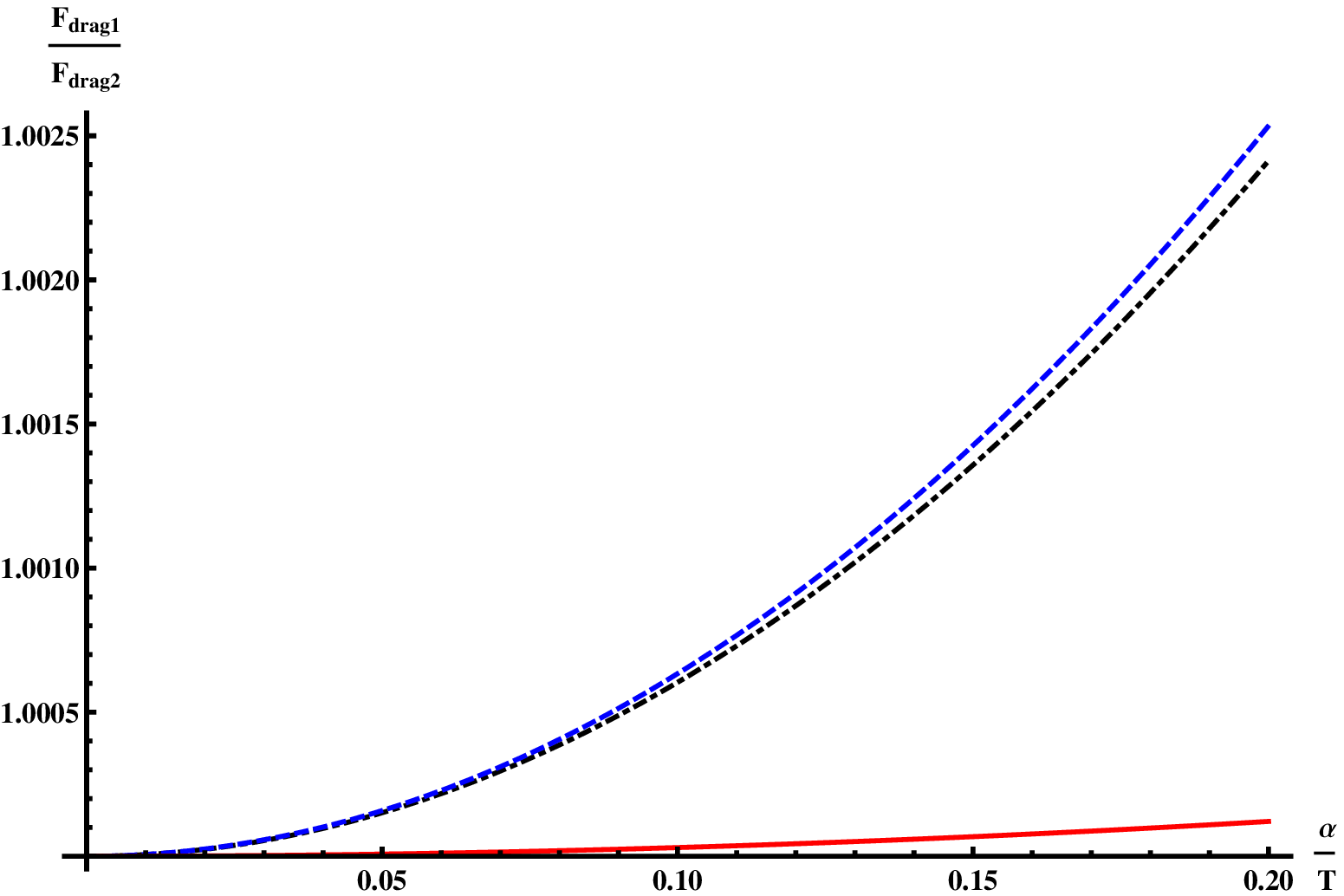}}
\caption{\small{The drag force dependence on the anisotropic parameter $a$. The velocity is chosen close to 1, $v\simeq 0.98$. The exact value of the velocity as far as is in this region, does not affect qualitatively the results. All the ratios in this plot are greater than one but for $v\lesssim 0.909$ the $F_{drag,\perp}/F_{drag,iso}$ is lower than unit and reduces as the anisotropy increases.
Settings: black dotdashed line-$F_{drag,\parallel}/F_{drag,\perp}$, blue dashed line-\blue{$F_{drag,\parallel}/F_{drag,iso}$}, red solid line-\red{$F_{drag,\perp}/F_{drag,iso}$} and $T=1$.
}}
\label{fig:figd1}
\end{flushleft}
\end{minipage}
\hspace{0.3cm}
\begin{minipage}[ht]{0.5\textwidth}
\begin{flushleft}
\centerline{\includegraphics[width=70mm]{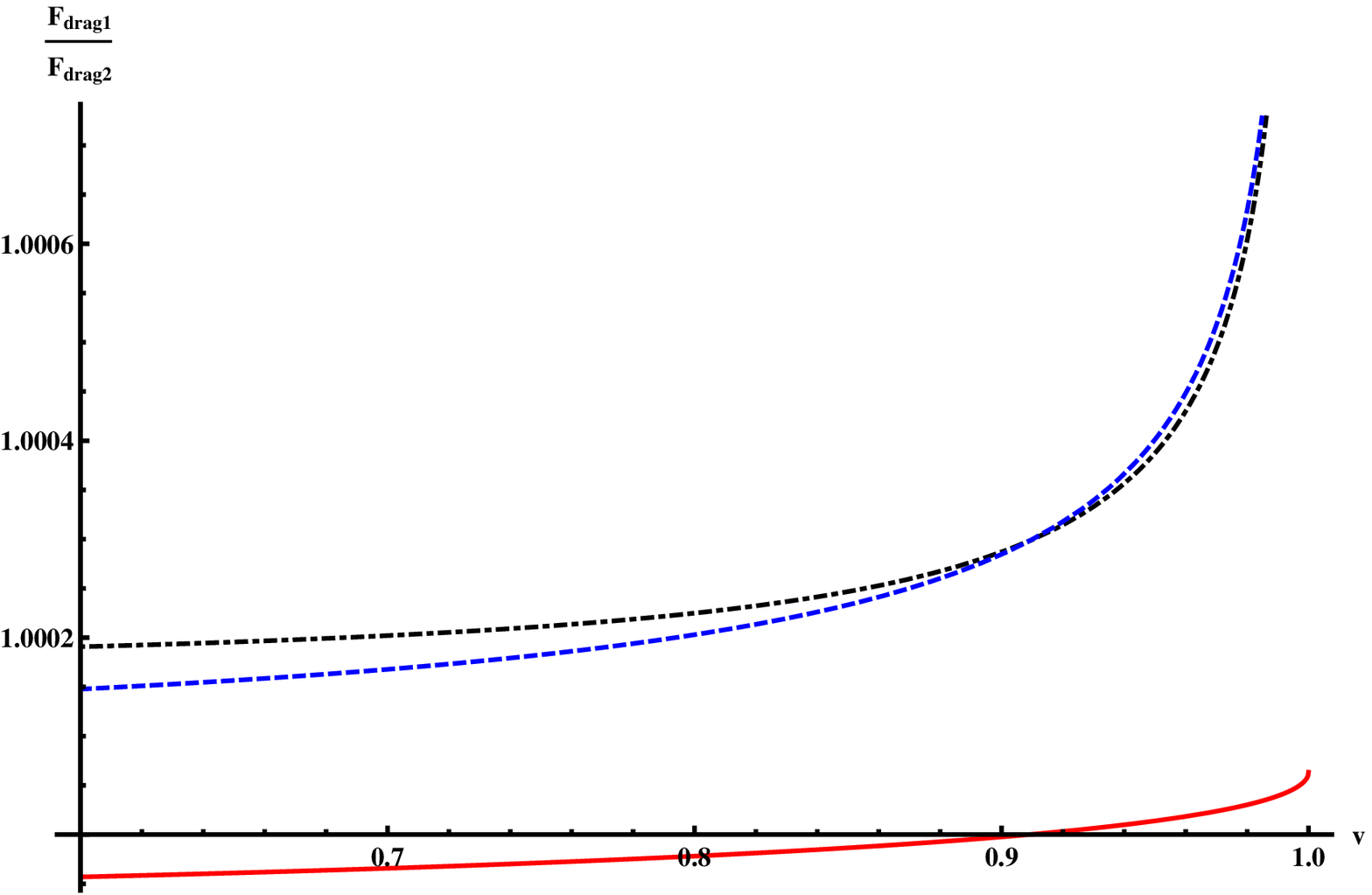}}
\caption{\small{The drag force dependence on the quark velocity $v$ for fixed $a$ and $T$. The only quantity lower that $1$ is the $F_{drag,\perp}/F_{drag,iso}$ for $v\lesssim 0.909$.
Settings: as in Figure \ref{fig:figd1} and  $a=0.1$ and $T=1$. \vspace{2.6cm} }}
\label{fig:figd2}
\end{flushleft}
\end{minipage}
\begin{minipage}[ht]{0.5\textwidth}
\begin{flushleft}
\centerline{\includegraphics[width=70mm]{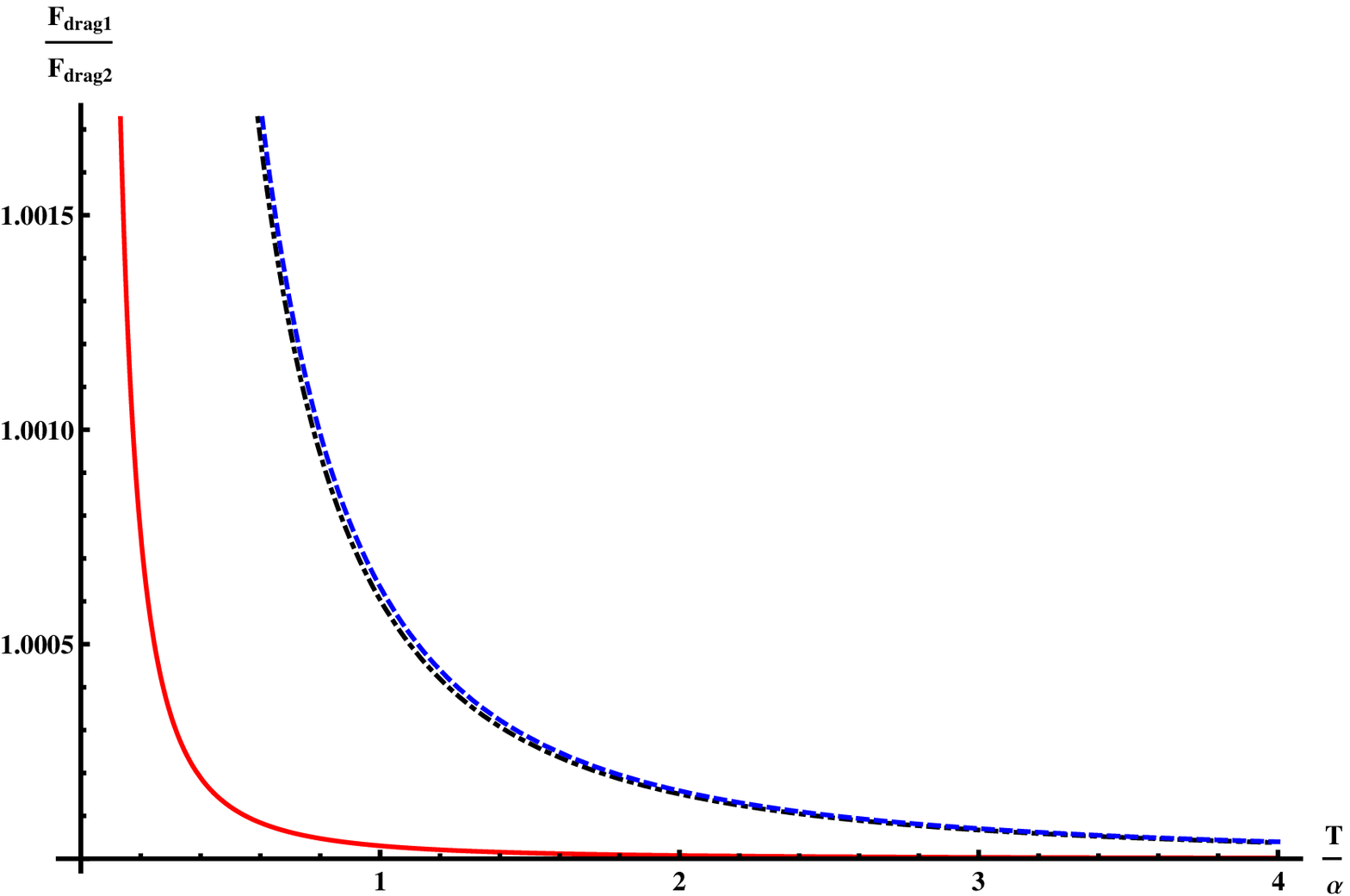}}
\caption{\small{The drag force ratios dependence on the temperature $T$. We choose $v=0.98$, where all ratios are above unit. As the temperature increases the ratios decrease.
Settings: as in Figure \ref{fig:figd1} and $a=0.1$.}}
\label{fig:figd3a}
\end{flushleft}
\end{minipage}\hspace{0.3cm}
\begin{minipage}[ht]{0.5\textwidth}
\begin{flushleft}
\centerline{\includegraphics[width=70mm]{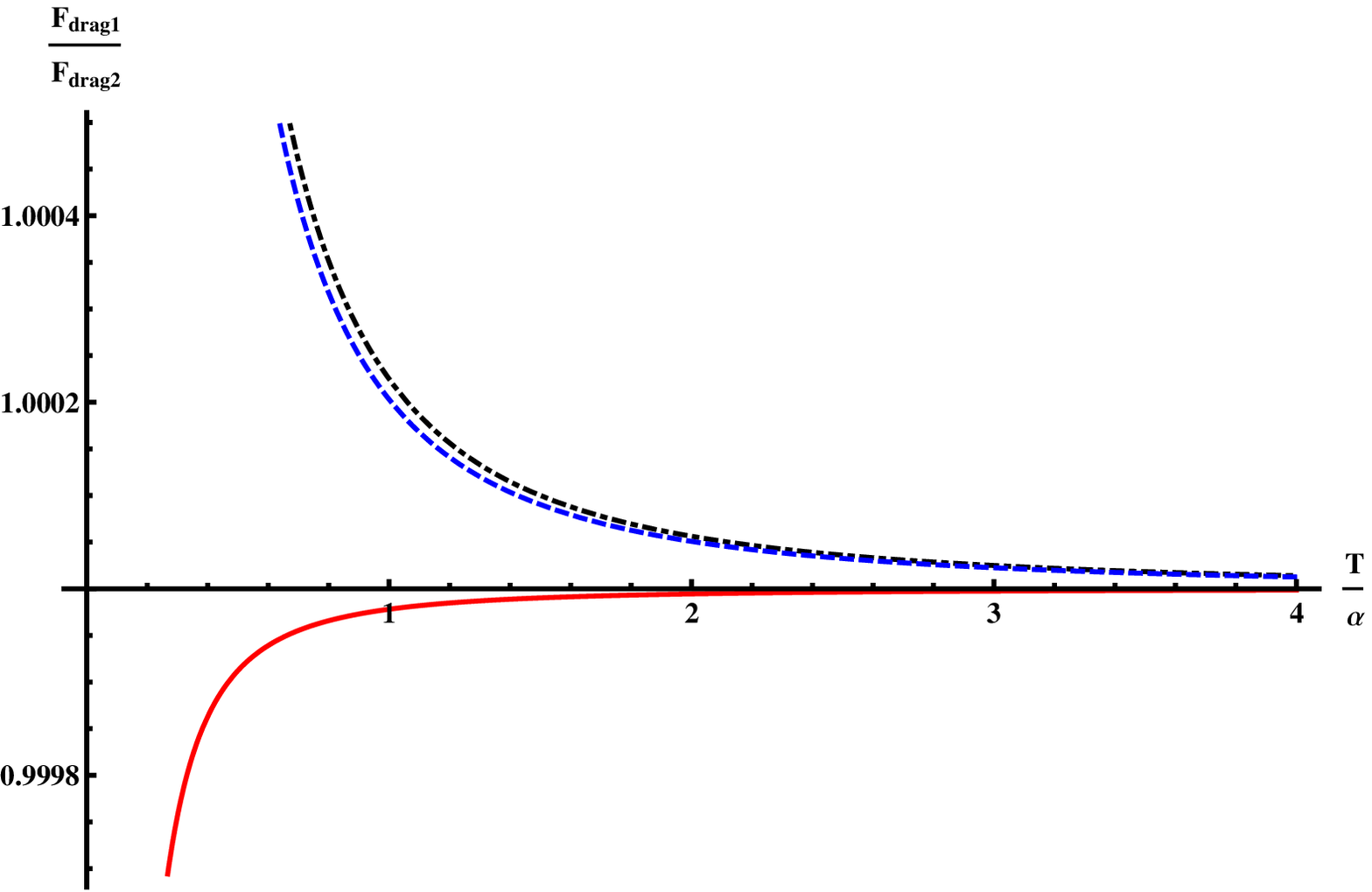}}
\caption{\small{The drag force ratios dependence on the temperature $T$. We choose $v=0.8$ where  the ratio $F_{drag,\perp}/F_{drag,iso}< 1$. As the temperature increases the deformed forces approach to the undeformed one.
Settings: as in Figure \ref{fig:figd1} and $a=0.1$.\vspace{-0.55cm}}}
\label{fig:figd3b}
\end{flushleft}
\end{minipage}
\end{figure*}

The $a^2$ term in the $F_{drag,\parallel}$ is always negative for any value of $v$ indicating enhanced drag force for motion along the anisotropic direction. The corresponding term in $F_{drag,\perp}$ is positive for velocities smaller than $v\leq v_c\simeq 0.909$ indicating decrease
of the drag force in this region, compared to the undeformed force. Above this critical value the drag force even in the transverse direction is enhanced, compared to the isotropic one as can be seen from  \eq{fper}. It is also worth noticing that the value of the critical velocity is independent of the temperature and anisotropy.

Moreover the drag force in the anisotropic direction is always greater than the one in the transverse direction and their ratio reads:
\be\label{fdpp}
\frac{F_{drag,\parallel}}{F_{drag,\perp}}=1+a^2 \frac{ \left(2-v^2\right) \log\left(1+\sqrt{1-v^2}\right)}{8 \pi ^2 T^2 \left(1-v^2\right)}~.
\ee
The correction term is always positive and depends on the temperature as expected from the dimensional analysis.

For better understanding of our findings we plot the following results.
In the  Figure \ref{fig:figd1} are presented the drag forces along parallel and transverse directions normalized with the undeformed isotropic result $F_{drag,iso}$, as well as the fraction $F_{drag,\parallel}/F_{drag,\perp}$, depending on the anisotropy. We see that increase of the anisotropy leads to increase of the deviation of all the drag forces. The strongest dependence on anisotropy is observed for quarks moving along the anisotropic direction as was also observed for the static potential.

In the Figure \ref{fig:figd2} we plot the same quantities depending on the velocity of the quark for constant $a$ and $T$. We observe that increase of the velocity of the probe quark leads to increase of the drag force. Where again for high enough velocities the drag force for motion along the anisotropic direction is affected stronger by the anisotropy. From the plot it is clear that for  $v\lesssim 0.909$ the drag force in an isotropic background is stronger than the force on the quark moving along the transverse direction to anisotropy. For higher velocities all the drag forces are enhanced compared to the isotropic case.

In Figures \ref{fig:figd3a}, \ref{fig:figd3b} we plot the drag force ratios for fixed anisotropy and velocities. We observe the differences for $v<v_c$ and $v>v_c$. Common in both plots is that as the temperature increases with respect to anisotropy the drag forces approach the isotropic ones as expected by construction of our background.

To summarize qualitatively our results:
\bea\nn
&&F_{drag,\parallel}>F_{drag,iso}\quad \mbox{and}\quad F_{drag,\parallel}>F_{drag,\perp}~,\\\nn
&&F_{drag,\perp}>F_{drag,iso}\quad \mbox{for}\quad v>v_c,\quad \mbox{while below this velocity}\quad F_{drag,\perp}<F_{drag,iso} .
\eea

\subsection{The drag coefficient and the diffusion time}

The drag coefficient is defined as
\be
\frac{dp}{dt}=-n_D p~,\quad\mbox{with}\quad p=\frac{M_Q v}{\sqrt{1-v^2}}~.
\ee
Therefore the diffusion time $\t_D$ is given by:
\be\label{timea1}
\t_{D,\parallel,\perp}=\frac{1}{n_{D,\parallel,\perp}}=-\frac{1}{F_{drag,\parallel,\perp}}
\frac{M_Q v}{\sqrt{1-v^2}}~
\ee
and corresponds to the time where the initial momentum is reduced by $e^{-1}$ factor. The equation \eq{timea1} indicates that the time inequalities between the different directions will be just inverted compared to the drag force results. For example the ratio of the diffusion time for motion along parallel and transverse direction is given by:
\be\label{tdpp}
\frac{\t_{D,\parallel}}{\t_{D,\perp}}=
1-a^2\frac{\left(2-v^2\right) \log\left(1+\sqrt{1-v^2}\right)}{8 \pi ^2 T^2 \left(1-v^2\right)}~,
\ee
\begin{figure*}[!ht]
\begin{minipage}[ht]{0.5\textwidth}
\begin{flushleft}
\centerline{\includegraphics[width=70mm]{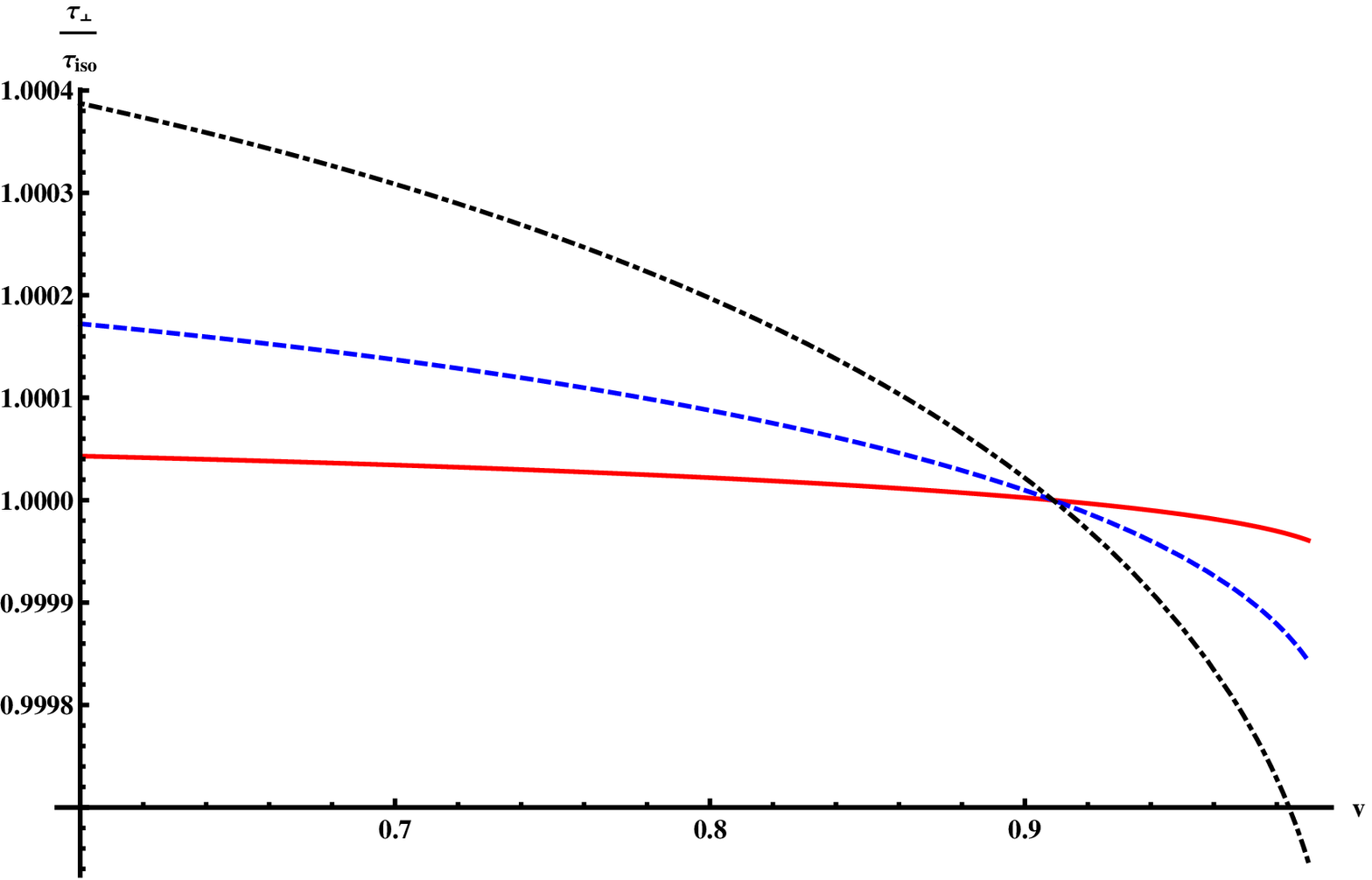}}
\caption{\small{The ratio of the diffusion time in the transverse direction of the anisotropic holographic model to the diffusion time of the isotropic theory.  For $v\lesssim 0.909$ the ratio is above $1$. 
Settings: red solid line-\red{$a=0.1$}, blue dashed line-\blue{$a=0.2$}, black dotdashed line-$a=0.3$ and $T=1$.
}}
\label{fig:fige1}
\end{flushleft}
\end{minipage}
\hspace{0.3cm}
\begin{minipage}[ht]{0.5\textwidth}
\begin{flushleft}
\centerline{\includegraphics[width=70mm]{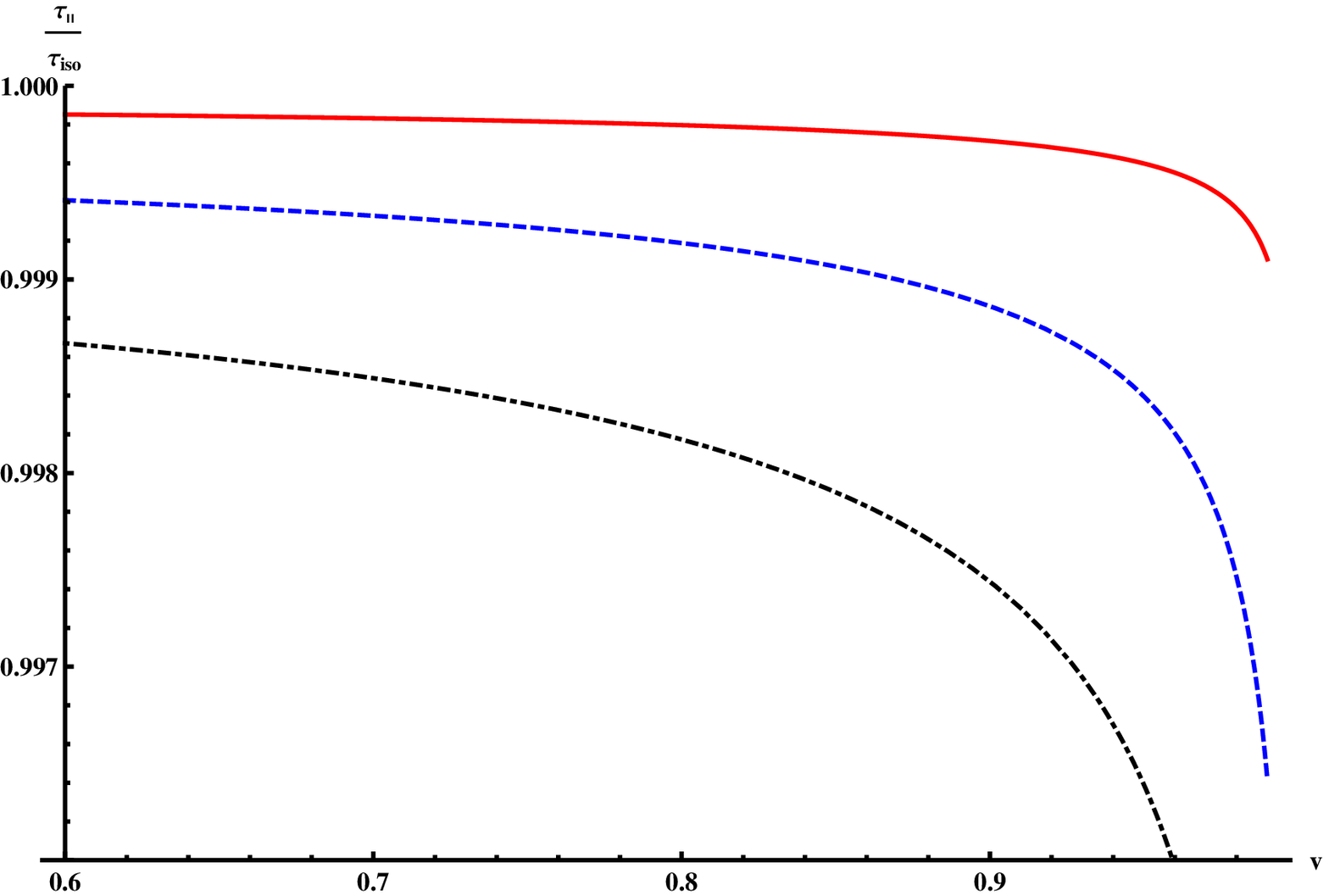}}
\caption{\small{The ratio of diffusion time for motion along the anisotropic direction to the diffusion time of the isotropic theory for different values of the anisotropic parameter. Settings as in Figure \ref{fig:fige1}.\vspace{1.0cm}}}
\label{fig:fige2}
\end{flushleft}
\end{minipage}
\begin{minipage}[ht]{0.51\textwidth}
\begin{flushleft}
\centerline{\includegraphics[width=70mm]{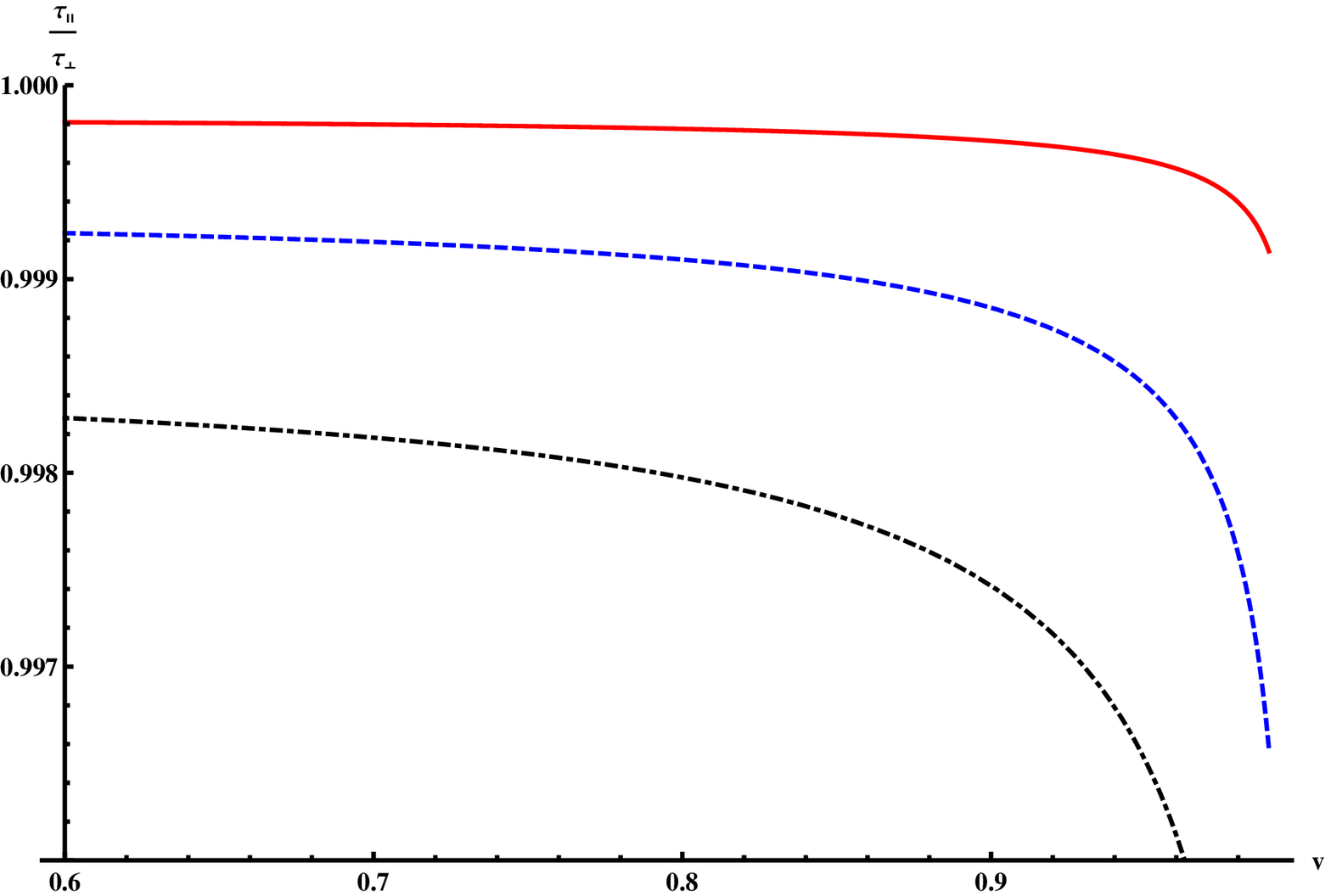}}
\caption{\small{The ratio of diffusion times along the anisotropic direction to the transverse one is always lower than the unit. For higher velocities the ratio diverges increasingly from the unit. Settings as in Figure \ref{fig:fige1}.
}}
\label{fig:fige3}
\end{flushleft}
\end{minipage}
\end{figure*}
meaning that the diffusion time along the anisotropic direction is lower compared to the one in the isotropic direction.
The other two ratios are\footnote{Notice that here we are considering as negligible the modifications to the mass of the quark from the thermal medium. As we discuss in section 6, the medium induced corrections to the mass are independent of the directions in anisotropic theory, but are expected to be slightly modified compared to the isotropic case.}:
\bea\label{tdpar}
\frac{\t_{D,\parallel}}{\t_{D,iso}}&=&1-\frac{a^2}{T^2}\frac{ \sqrt{1-v^2}\left(1+\sqrt{1-v^2}\right)+\left(1+v^2\right) \log\left(1+\sqrt{1-v^2}\right)}{24 \pi ^2 \left(1-v^2\right)}~,
\\\label{tdper}
\frac{\t_{D,\perp}}{\t_{D,iso}}&=&1-\frac{a^2}{T^2}\frac{\sqrt{1-v^2}\left(1+\sqrt{1-v^2}\right)- \left(5-4 v^2\right) \log\left(1+\sqrt{1-v^2}\right)}{24 \pi ^2 \left(1-v^2\right)}~.
\eea
The diffusion time $\t_{D,\perp}$ is longer than the isotropic one for $v< v_c$, while for $v>v_c$ it is shorter as expected from the drag force results.
In Figures \ref{fig:fige1}, \ref{fig:fige2}, \ref{fig:fige3} we plot the ratios of diffusion times with respect to velocity of the quarks for three different values of $a$.  Increase of the velocity of the moving quark or increase of the anisotropy makes the ratios $\t_{D,\parallel}/\t_{D,(iso,\perp)}$ to increasingly diverge from the unit.

To briefly summarize, we find that for very high velocities the diffusion time in presence of anisotropy is reduced compared to the isotropic theory. Hence the diffusion process is faster in presence of the anisotropy for high velocities.  We also find that the anisotropy affects more the diffusion time in the anisotropic direction that in the transverse one. More particularly the relation between the different diffusion times are:
\bea
&&\t_\parallel<\t_{iso} \quad \mbox{and}\quad\t_\parallel<\t_\perp~,\\
&&\t_\perp<\t_{iso}\quad \mbox{for}\quad v>v_c,\quad \mbox{while below this velocity}\quad \t_\perp>\t_{iso}~.
\eea

\section{Jet Quenching in the anisotropic strongly coupled $\cN=4$ sYM plasma}

In this section we calculate the anisotropy effects on the jet quenching parameter. Its bounds can be measured in the QGP by the radiative energy loss and the parameter itself can be though as a property of the strongly coupled medium. The jet quenching is generated when the momentum of an energetic parton changes while it moves in the medium. The interaction with the medium result the parton to radiate gluons which is the reason of the transverse momentum broadening. The jet quenching  is defined as the fraction of the mean transverse momentum obtained by the hard parton in the medium over the distance it has traveled.

In the field theory the transverse momentum broadening can be calculated using a Wilson line in the adjoint representation along a light cone direction. In the gravity dual description the jet quenching can be calculated from the minimal surface of a world-sheet which ends on an orthogonal Wilson loop lying along two light-like lines. These two long parallel lines of the Wilson loop, with length say $L_{-}$, are related to the partons moving at relativistic velocities and are taken to be much more larger that the other two sides of the loop with length $L_k$ related to the transverse momentum of the radiated gluons.

The Wilson loop we calculate in supergravity side is in fundamental representation but in the planar limit the expectation value of the adjoint Wilson loop is related to the Wilson loop in the fundamental representation as
\be\label{waf}
\vev{W^A(\cC)}=\vev{W^F(\cC)}^2~,
\ee
because $tr_{Adj}=tr^2_{Fund}$. The LHS of the above equation is related to the jet quenching parameter as \cite{jetq}
\be
\vev{W^A(\cC)}\approx\exp^{-\frac{1}{4\sqrt{2}}\hat{q}L_{k}^2 L_-}~.
\ee
To calculate the corresponding Wilson loop we go to the light-cone coordinates by the coordinate transformation $\sqrt{2} x^\pm=t \pm x_p$, where $x_p$ is chosen to be $x_{\parallel}$  or $x_\perp$, as in  \eq{xppp}~.
The generic metric \eq{metricqq} becomes
\bea
&&ds^2=G_{--} (dx_{+}^2+ dx_{-}^2)+G_{+-} dx_+ dx_- +G_{ii(i\neq p)}dx_i^2+G_{uu} du^2\\\nn
&&G_{--}=\frac{1}{2}(G_{00}+G_{pp}),\qquad G_{+-}=G_{00}-G_{pp}
\eea
Taking advantage of the condition $L_- \gg L_k$, where $L_k$ is the length of the string in the $k$ direction, we assume that the string worldsheet is translational invariant along  $x_k$. This simplifies things significantly. We present here the full calculation of the
jet quenching because under certain approximations we can arrive in an analytic expression valid for any background. We numerically check our analytical expressions with the exact ones and we find that our approximations are correctly considered.

The ansatz for the string configuration is
\bea
&&x_-=\tau,\quad x_k=\sigma,\quad u=u(\s) \\
&&x_+,~x_{i\neq p}\quad \mbox{are constant}~,
\eea
which represents a Wilson loop extending along the $x_k$ direction and lying at a constant $x_+, x_{i\neq p}$. The indices $p,~k$ here denote a chosen direction. The action then reads
\be
S=\frac{2L_-}{2 \pi a'}\int_0^\frac{L_k}{2} d\s \sqrt{G_{--}(G_{uu}u'^2+G_{kk})}=:\frac{2L_-}{2 \pi a'}\int_0^\frac{L_k}{2} d\s \sqrt{D}
\ee
and the analysis has some similarities to the Q\={Q} action analysis presented Appendix A. Using Hamiltonian formalism we obtain
\be
H=\frac{G_{--}G_{kk}}{\sqrt{D}}~.
\ee
The Hamiltonian is a constant of motion, and we set it equal to $c$. We can solve for the $u'$ and get
\be
u'^2=\frac{(G_{kk} G_{--}-c^2)G_{kk}}{c^2 G_{uu}}~,
\ee
where at the turning point the string satisfies
\be
G_{kk}=0~,\quad\mbox{or}\quad G_{kk} G_{--}=c^2,\quad\mbox{or}\quad G_{uu}^{-1}=0~.
\ee
Usually the interesting equation in this case is the last one which is satisfied for $u=u_0=u_h$. The short length of the string is then given by
\be
\frac{L_k}{2}=\int_{0}^{u_h} du\sqrt{\frac{c^2 G_{uu}}{(G_{kk}G_{--}-c^2)G_{kk}}}~.
\ee
Since we are interested in the small $L_k$ length and the integral goes from the boundary to the turning point, the constant $c$ must be very small. Therefore, we expand our formula, and at the end we will check the validity of the expansion numerically to justify it. The constant of motion turn out to be given by the analytic expression
\be
c=\frac{L_k}{2}\left(\int_{0}^{u_h} du \frac{1}{G_{kk}}\sqrt{\frac{G_{uu}}{G_{--}}}\right)^{-1}+ \mathcal{O}(L_k^3)~.
\ee
Before we substitute to the action we need to eliminate the infinity in the action that appears due to the bounds of the integral. We use
the mass subtraction scheme where we subtract the two straight string
world sheets described by: $x_-=\t, u=\s$. The self energy reads:
\be
S_0=\frac{2 L_-}{2 \pi \a'}\int_{0}^{u_h} du \sqrt{G_{--}G_{uu}}~.
\ee
The total action subtracted the divergences is equal to:
\be\label{sss0}
S-S_0=\frac{2 L_-}{2 \pi \a'}\int_{0}^{u_h} du \sqrt{G_{uu}G_{--}}\left(\sqrt{\frac{G_{--}G_{kk}}{G_{--}G_{kk}-c^2}}-1\right)~.
\ee
After some algebra the normalized action takes the simple form
\be
S-S_0=\frac{L_- L_k}{4\pi a'}c~,
\ee
where expressed  in metric elements
\be\label{sjet}
S=\frac{L_- L_k^2}{8\pi a'}\left(\int_{0}^{u_h} du \frac{1}{G_{kk}}\sqrt{\frac{G_{uu}}{G_{--}}}\right)^{-1} +\mathcal{O}(L_k^4)~.
\ee
We have checked numerically that in our case the small $c$ approximation is valid where the equations \eq{sss0} and \eq{sjet} take the same values for small values of $c$.
The jet quenching parameter for an energetic parton moving along the $p$ direction while the broadening happens along the $k$ direction is given by:
\be\label{qfinal}
\hat{q}_{p (k)}=\frac{\sqrt{2}}{\pi\a'}\left(\int_{0}^{u_h} \frac{1}{G_{kk}}\sqrt{\frac{G_{uu}}{G_{--}}}\right)^{-1}~,
\ee
where to normalize correctly we have taken into account the equation \eq{waf}. The equation \eq{qfinal} is very useful, since can be applied directly to any background that satisfies the approximations we have made here.

For our 4-dim anisotropic plasma we have three different choices for the transverse momentum broadening. The first one, $\hat{q}_{\parallel(\perp)}$, is for the energetic parton moving parallel to the anisotropic direction and the momentum broadening occurring along the transverse direction. The second one, $\hat{q}_{\perp(\parallel)}$, is for an energetic parton moving along $x_1$ or $x_2$ direction and the momentum broadening happens along the anisotropic $x_3$ direction. The last one, $\hat{q}_{\perp(\perp)}$, is for a parton moving along the transverse to the anisotropy directions and the momentum broadening is considered along the other parallel direction. In this case although only the directions along the transverse space are involved, the dependence of the radial direction metric element on the anisotropic parameter, modify although sightly the result compared to the isotropic theory.
Therefore the configurations we consider are:\newline\newline
\begin{tabular}{|c|c|c|c|c|}
\hline
$\hat{q}$&$x_p$ & $x_k$ & Energetic parton moves along& Momentum broadening along\\\hline
$\hat{q}_{\perp(\parallel)}$&$x_\perp$ &$x_\parallel$& $x_\perp$& $x_\parallel$ \\ \hline
$\hat{q}_{\parallel(\perp)}$&$x_\parallel$ &$x_\perp$& $x_\parallel$& $x_\perp$\\
\hline
$\hat{q}_{\perp(\perp)}$&$x_{\perp,1}$ &$x_{\perp,2}$& $x_{\perp,1}$& $x_{\perp,2}$ \\
\hline
\end{tabular}\newline\newline
Where the $x_{\parallel,\perp}$ are defined with respect to the anisotropy direction.

\begin{figure*}[!ht]
\begin{minipage}[ht]{0.5\textwidth}
\begin{flushleft}
\centerline{\includegraphics[width=70mm]{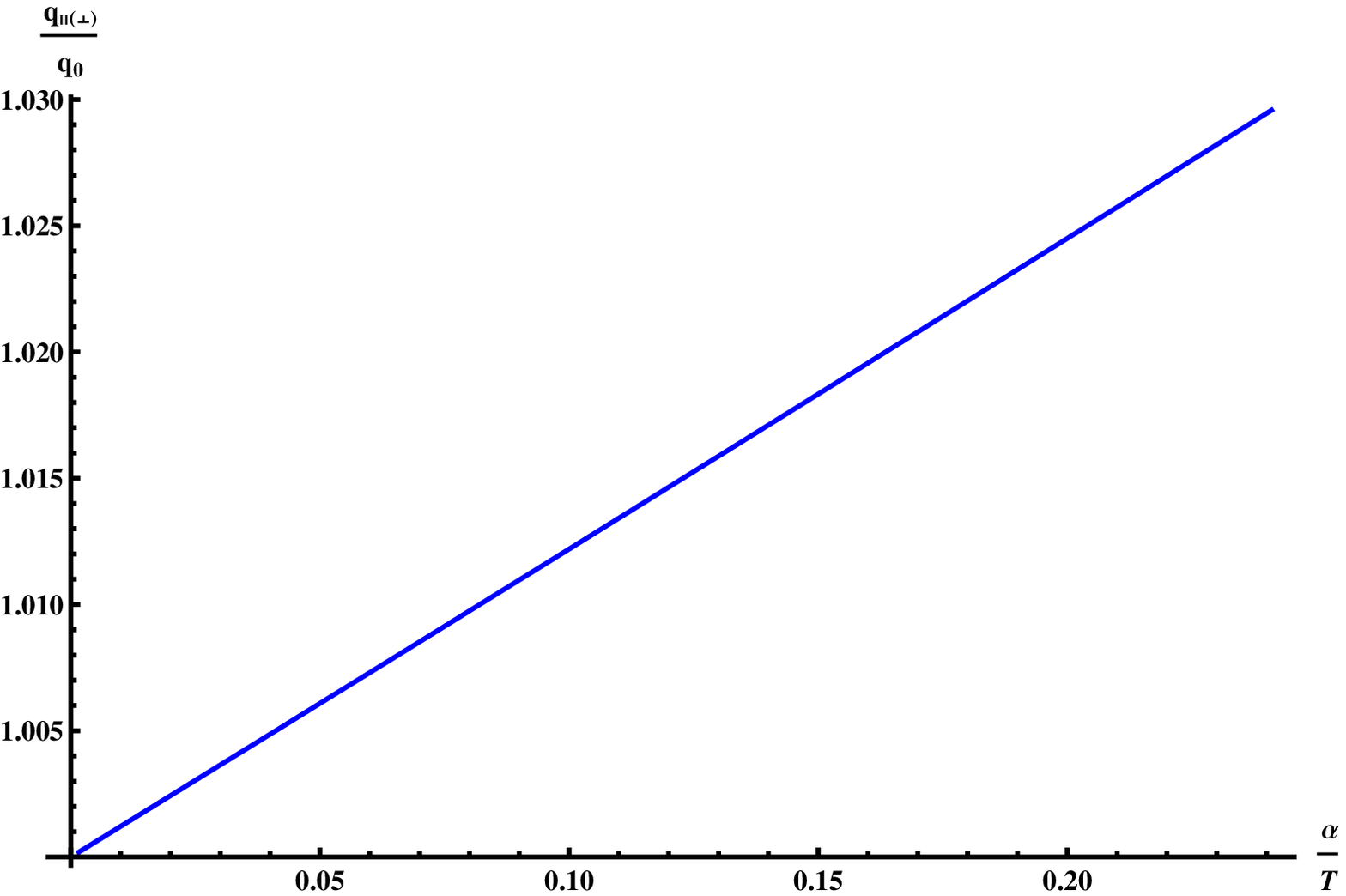}}
\caption{\small{The  jet quenching for a heavy quark moving along the anisotropic direction $\hat{q}_{\parallel(\perp)}$ as a function of $a/T$. The results are normalized with the isotropic jet quenching. Notice that normalization with $\hat{q}_{\perp(\perp)}$ gives almost identical results.
Settings:  $T=5$.}}
\label{fig:figf1}
\end{flushleft}
\end{minipage}
\hspace{0.3cm}
\begin{minipage}{0.5\textwidth}
\begin{flushleft}
\centerline{\includegraphics[width=70mm]{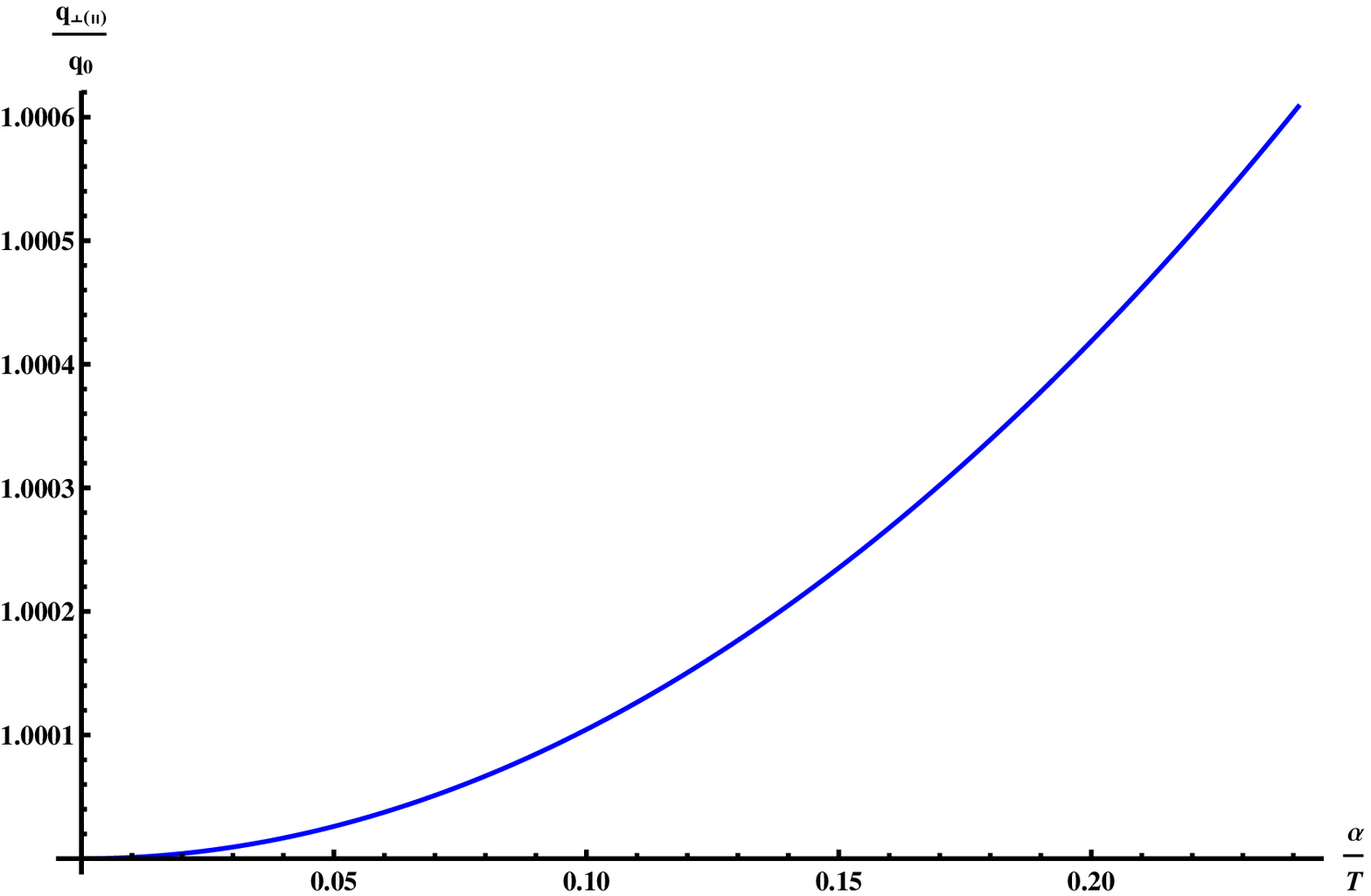}}
\caption{\small{The momentum broadening $\hat{q}_{\perp(\parallel)}$ along the anisotropic direction for a heavy quark moving in the transverse space.
Settings: $T=5$.\vspace{0.9cm}}}
\label{fig:figf2}
\end{flushleft}
\end{minipage}
\begin{minipage}[ht]{0.5\textwidth}
\begin{flushleft}
\centerline{\includegraphics[width=70mm]{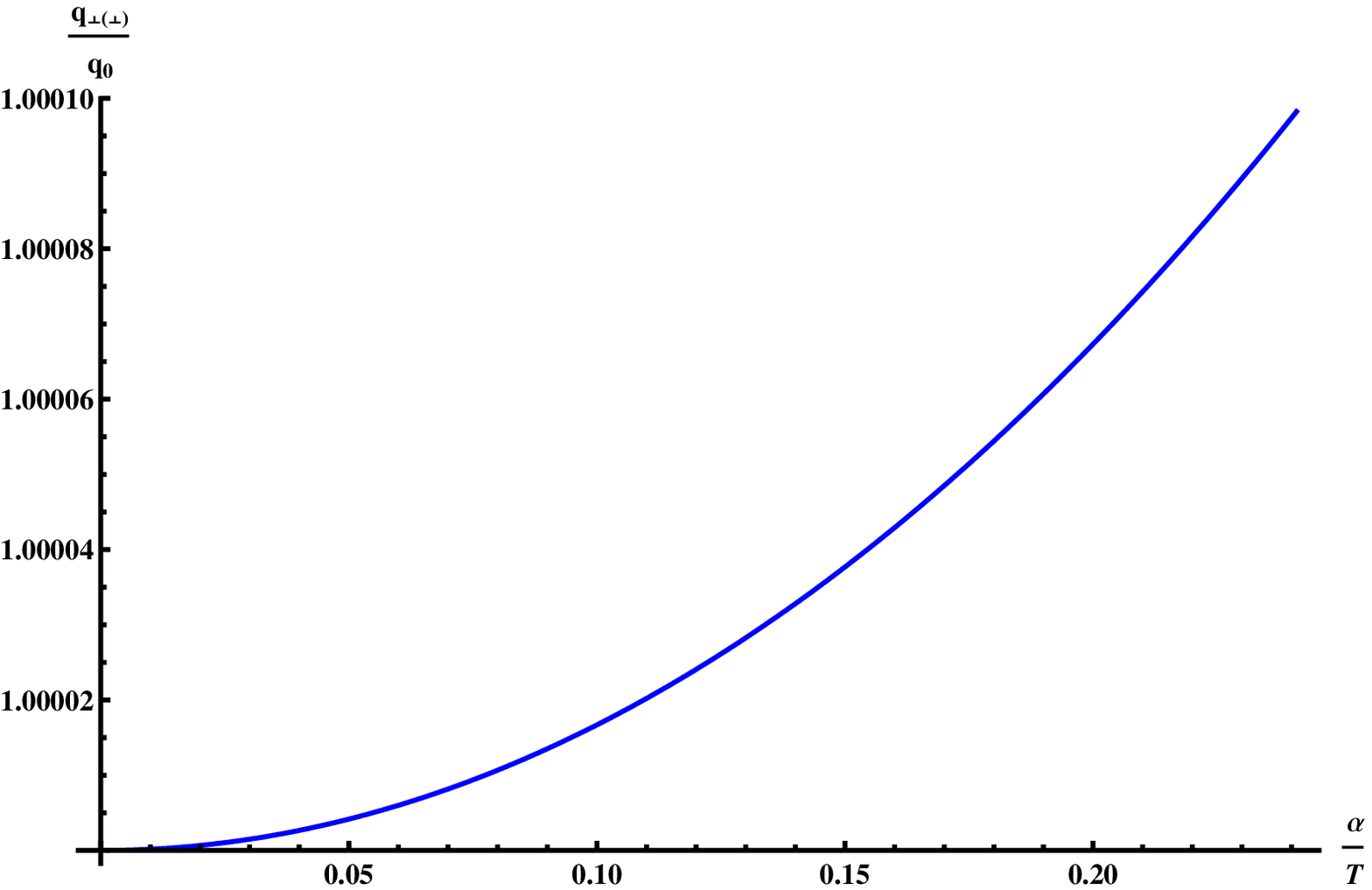}}
\caption{\small{Quark motion and broadening in transverse space:$\hat{q}_{\perp(\perp)}$.
Settings: $T=5$.\vspace{1.7cm}}}
\label{fig:figf3}
\end{flushleft}
\end{minipage}\hspace{0.3cm}
\begin{minipage}[ht]{0.5\textwidth}
\begin{flushleft}
\centerline{\includegraphics[width=70mm]{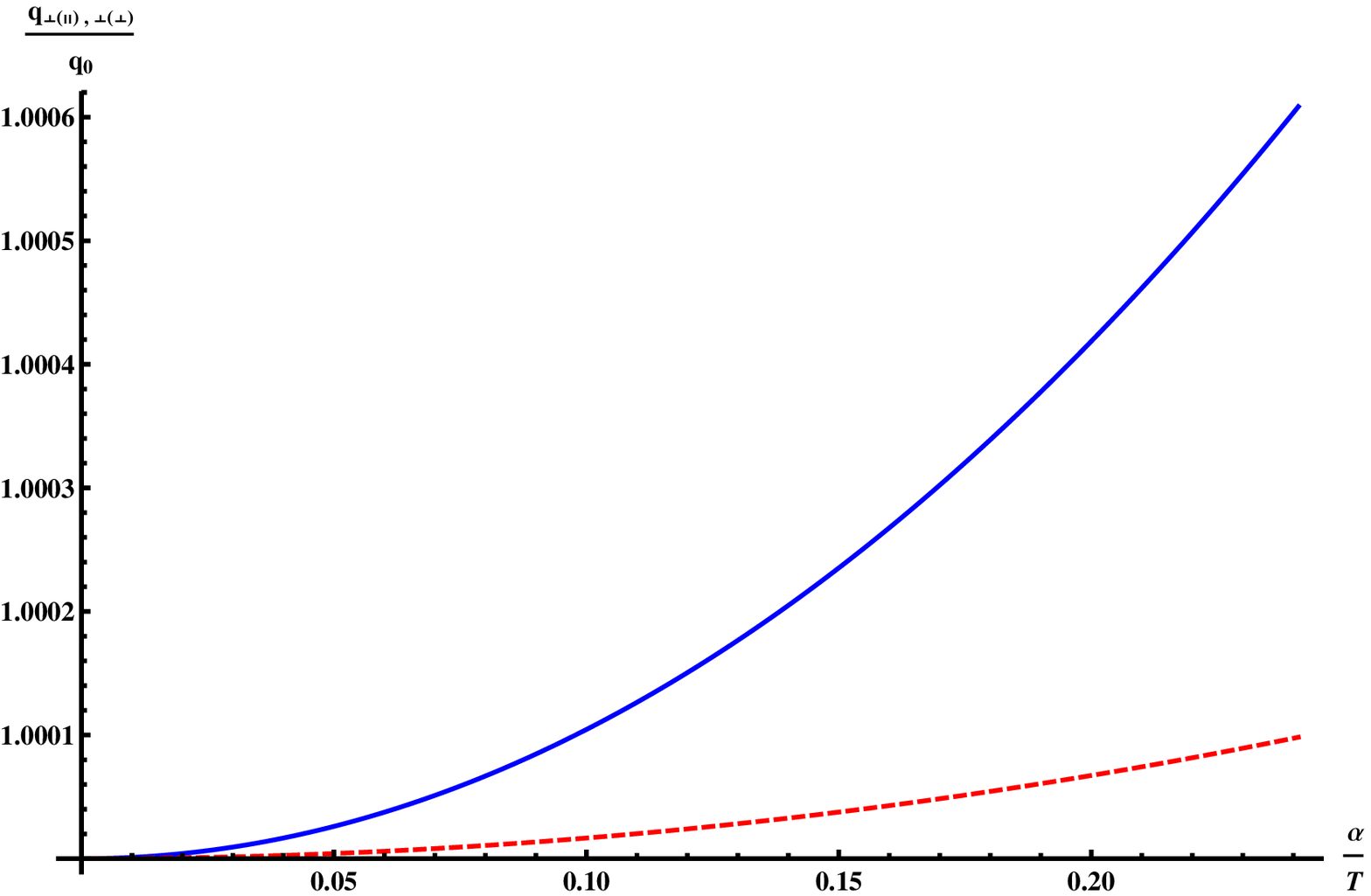}}
\caption{\small{Comparing the jet quenching parameters. The $\hat{q}_{\parallel(\perp)}$ is much more bigger than the other two as expected. The $\hat{q}_{\perp(\parallel)}$, $\hat{q}_{\perp(\perp)}$ are closer and are plotted here.
Settings: Blue solid line-\blue{$\hat{q}_{\perp(\parallel)}$}, red dashed line-\red{$\hat{q}_{\perp(\perp)}$},  $T=5$.}}
\label{fig:figf4}
\end{flushleft}
\end{minipage}
\end{figure*}
Let us start with the case where the quark moves along the anisotropic direction and the momentum broadening occurs along the transverse one. Then
\be
G_{pp}=G_{33}\qquad\mbox{and}\qquad G_{kk}=G_{11}~.
\ee
We find that the jet quenching in presence of anisotropy is enhanced with respect to the isotropic case. Stronger anisotropy leads to further enhancement of the fraction $\hat{q}_{\parallel(\perp)}/\hat{q}_{0}$, where $\hat{q}_{0}:=\hat{q}_{iso}$. Notice that normalization with $\hat{q}_{\perp(\perp)}$ instead of $\hat{q}_{0}$, leads to almost identical results. In Figure \ref{fig:figf1} we show some representative results. The form of the function is linear and particularly:
\be
\frac{\hat{q}_{\parallel(\perp)}}{\hat{q}_{0}}\simeq 1+ 0.122 \frac{a}{T}
\ee
in a very good approximation, where the numerical factor does not depend strongly on the other parameters of our background. However, it is normal to expect that the lower order in $a/T$ will be $(a/T)^2$ and the above linear behavior appears only approximately in small anisotropies. The form of the function then becomes:
\be
\frac{\hat{q}_{\parallel(\perp)}}{\hat{q}_{0}}\simeq 1+ 0.475 \left(\frac{a}{T}\right)^2
\ee
Comparing to the other jet quenching parameters $\hat{q}_{\perp(\parallel)},~ \hat{q}_{\perp(\perp)}$ we will see that the anisotropy affects stronger the jet quenching for the quark moving along the anisotropic direction than the other ones.

To examine the $\hat{q}_{\perp(\parallel)}$ we need to set
\be
G_{pp}=G_{11}\qquad\mbox{and}\qquad G_{kk}=G_{33}
\ee
and substitute to \eq{qfinal}. Some representative results are
shown in the Figure \ref{fig:figf2}. Again we observe enhancement of the jet quenching, but this time is much more weaker than the $\hat{q}_{\parallel(\perp)}$. The form of the function is no more approaching to linear behavior, but increase of anisotropy still leads to increase of the jet quenching.

Finally, we look at $\hat{q}_{\perp(\perp)}$ where we need to choose
\be
G_{pp}=G_{11}\qquad\mbox{and}\qquad G_{kk}=G_{22}~.
\ee
We show that a weak enhancement of the jet quenching observed as can be also seen in Figure \ref{fig:figf3}. Although the quark motion and the
momentum broadening happen along the transverse directions the jet quenching gets modified due to the fact that the time and the radial metric elements are dependent on the anisotropy parameter. We compare $\hat{q}_{\perp(\perp)}$ with $\hat{q}_{\perp(\parallel)}$ in Figure \ref{fig:figf4} where we observe important differences in their magnitudes.

There are some common features that we observe for all the jet quenching parameters. We find that when anisotropy is present the jet quenching is enhanced. Increasing the anisotropy parameter leads to increase of the enhancement. The presence of anisotropy affects mostly the momentum broadening of a quark moving along the anisotropic direction. This is something we have seen in the static potential and the drag force. Alignment of the Q\={Q} pair along the anisotropic direction, as well as motion of the quark with the trailing string along this direction led to maximum modifications on the relevant results due to anisotropies.

To summarize our findings: We find that the jet quenching is in generally enhanced in presence of anisotropy compared to the isotropic case and that its value depends strongly on the direction of the moving quark and the direction which the momentum broadening occurs. More particularly
\be
\hat{q}_{\parallel(\perp)}>\hat{q}_{\perp(\parallel)}>\hat{q}_{\perp(\perp)}>\hat{q}_{iso}~,
\ee
which in our conventions translated as: the jet quenching is stronger enhanced for a quark moving along the anisotropic direction and momentum broadening happens along the transverse one. The parameter gets lower for a quark moving along the transverse direction and the momentum broadening considered along the anisotropic one. A very weak enhancement is observed when the quark moves in the transverse plane and the broadening happens on the same plane.

\subsection{Comparison with other results}

It has been shown in many different studies that  certain jet quenching parameters in
an anisotropic plasma are increasing in presence of anisotropy. Moreover, in these studies the ordering of
$\hat{q}_{\perp(\parallel)},~\hat{q}_{\perp(\perp)}$ agrees with the one we have found here.
However, we should keep in mind that the models mentioned below have several other differences from our theory
 apart from the fact that are in the weak coupling limit. Nevertheless, the comparison is useful since
these differences are not a priori known how much they affect the corresponding results.

It has been found using kinetic theory that in the leading-log order the jet quenching of a heavy quark moving along one of the transverse directions to anisotropy is larger when the momentum broadening is along the anisotropic axis, which in this case coincide with the beam axis, compared to when the broadening occurs in the other transverse direction \cite{jetqa2}. In our conventions these results can be written as  $\hat{q}_{\perp(\parallel)}>\hat{q}_{\perp(\perp)}$. This is in agreement with our results as can be seen in Figure \ref{fig:figf4}.

In \cite{jetsu2} the jet quenching was calculated in an unstable non-Abelian weakly coupled $SU(2)$ plasma. Using numerical simulations
taking into account hard elastic collisions and soft interactions mediated by classical Yang-Mills fields, and a separation scale between them, the authors found that the fields develop unstable modes which lead to $\hat{q}_{\perp(\parallel)}>\hat{q}_{\perp(\perp)}$ in our notation.

In \cite{jetqa1} the jet quenching is estimated in leading logarithmic approximation by the broadening of the massless quark interacting via gluon exchange. The energetic hard quark considered propagates in one of the transverse directions to the anisotropy and the momentum broadening is estimated in the transverse to the motion plane, which include the anisotropic direction. The jet quenching in the anisotropic plasma was found to be enhanced with respect to the isotropic one.

It is important to notice here that our jet quenching results are in some agreement with STAR findings, eg. \cite{star}. This might be an additional indication that the QGP in LHC and RHIC is not in equilibrium.

\section{Attempts for more Quantitative Comparison}

In this section we make an attempt to predict the effect of the anisotropy on the observables in a more quantitative way. To do that we need to use sensible comparison schemes and give realistic values to the parameters of the model. There are several differences between our anisotropic deformed theory and the QCD but a comparison between strongly coupled plasmas in the two theories could be done under some logical normalization schemes.
For example one motivation for these comparisons is that there are indications from lattice QCD calculations which show that lattice QCD thermodynamics are in good approximation with conformal theories for some finite range of temperatures $T>2T_c$.

For the heavy test quarks we used we can choose the charm and bottom quarks. Their effective masses are difficult to be determined precisely in the thermal medium, but the most representative values are: $M_c=1.5 GeV$ and $M_b=4.8 GeV$. We can include the medium induced correction to these masses, by specifying their dependence on the temperature. As we mention in the Appendix A, the mass of the static quark is represented by the static straight string stretched along the radial radiation starting from the boundary of the space and reaching the black hole horizon. However, this string is infinite and that is the reason that is used to normalize the infinite static potential. Therefore in the UV, a regulator should be introduced to make the result finite. The value of the regulator has to be fixed by matching the $M_Q(T=0)$ to the physical quark mass. Notice that the medium induced corrections to the  masses are not affected of the direction of the anisotropy. On the other hand, they are very slightly modified compared to the isotropic case, since the horizon position is modified.

To compare the observables in the theories we need to fix appropriately the theory parameters. Our model is not a confining and the static potential does not include the linear term. However, by comparing the renormalized charge or the static force of the conformal theory with the lattice data for relatively small separation lengths of the quark pair, it is possible to find that a well estimated value in order to agree for the t' Hooft coupling is $\l=5.5$.
The next parameter we need to determine for comparison is the choice of temperature in our anisotropic deformed theory and the QCD. The degrees of freedom in two theories as well as the field content are very different. For example in our case we do not have flavors and we have very large number of the color branes. To qualitatively compare the two theories we normalize the quantities firstly according to a fixed energy density scheme. By approximating the QCD plasma as a free gas the energy density is
\be
\e_{QGP}\simeq\frac{\pi^2(N_c^2-1+N_c N_f)}{15}T_{QGP}^4\simeq 11.2 ~ T_{QGP}^4, \quad\mbox{where}\quad N_c=N_f=3~,
\ee
and the factor $N_c^2-1+N_c N_f$ is the degrees of freedom of $SU(N_c)$ QCD with $N_f$ flavors. By counting the degrees of freedom in $\N=4$ sYM can be found that there are approximately $2.7$ times more than the ones in QCD above the confinement phase transition, in our case $\simeq 45$.
Taking advantage of the fact that our theory is  deformed with a very small deformation parameter $a$ and equating the energy densities in these two theories we obtain the relation \cite{0902.4041}
\be
T_{SYM, anisot}=2.7^{-1/4}T_{QCD}~.
\ee
The next parameter we need to fix is the anisotropy.
From \eq{xia} we find that
\be\label{axi2}
a^2=\frac{8 \pi^2 T^2 \xi}{5}~
\ee
and therefore we are in the region of small $\xi$. If the anisotropic parameter $\xi$ is small then it is also related to the shear viscosity of the plasma, where in one-dimensional boost invariant expansion governed by Navier-Stokes evolution\cite{0902.3834,0608270,1007.0889,1101.4651}, the result given from \eq{xieta}
\be
\xi=\frac{10 \eta}{T\t s}
\ee
where the $\eta/s$ is the viscosity to entropy ratio, and $\t^{-1}$ defines the expansion rate, since $\t$ is the proper time. The dimensionless parameter $T\t$ determines the magnitude of the anisotropies and in the strong coupling in the RHIC and LHC initial conditions respectively this can be approximated to 
\bea
T\t\simeq0.35 \quad\mbox{for RHIC conditions}\\
T\t\simeq0.43 \quad\mbox{for LHC conditions}~,\label{Tt}
\eea
where we have assumed $\t_0\simeq 0.2 fm$, $T=350 MeV$ and $\t_0\simeq  0.1 fm$, $T=850 MeV$ respectively, since at LHC energies the initial time  is expected to be small.
To get a better picture of the above assumptions and the qualitative picture we fix the total entropy and observe the behavior of $T\t$. By assuming that the entropy scales as $s\propto T^3$ and $V_0\propto\t_0$, with $S=s_0 V_0$ we can estimate the corresponding times at different conditions. For example, by considering $T_0=250$ MeV to correspond to $\t_0\simeq 1$, then for temperature $T_0=350$ MeV the corresponding time is $\t_0\simeq 0.36$. Therefore, increase of temperature leads to decrease of $\t_0$ which is in agreement with the assumptions made to get \eq{Tt}.
By choosing in the strong coupling limit that $\eta/s\simeq 0.1$ as a representative value, we get for the initial conditions 
$\xi\simeq2.3$ and $\xi\simeq 2.8$,
where the smaller momentum space anisotropies correspond to the LHC initial conditions. Notice that reduction of the number of the flavor degrees of freedom, result also lower values for $\t_0$ since the temperature increases.

For RHIC energies the typical formation times cover a range of 
$0.2fm$ to less than $1 fm$ for temperatures $T_0=250-350$ MeV, estimation made by fixing the total entropy to reproduce the measured rapidity density of hadrons at a given centrality, e.g., at impact parameter $b \simeq7 fm$ \cite{qgpbook}. Additionally, the hydrodynamical models normally require thermalization times $\t_{therm}$ in the range of $0.6-1 fm$ in order to reproduce the magnitude of elliptic flow which is observed at RHIC. We need to choose a representative value for $\xi$, at $T=250$ MeV and by considering $\t\simeq 0.6 fm$, then $T \t\simeq 0.76$ and
\be\label{xirep}
\xi_R \simeq 1.30~.
\ee
This is the value we choose as a representative one for RHIC energy. Notice that the choice of the value depends strongly on the values we chose for the viscosity over entropy ratio and for the proper time. Since the exact values for both of these quantities are quite uncertain our numerical value should be taken with caution. For the viscosity over entropy ratio the value that is predicted in the AdS/CFT is around the well known $\eta/s\simeq 1/4 \pi$ \cite{Policastro:2001yc}. For RHIC conditions it has been estimated for a hot gluon plasma for $SU(3)$ pure gauge model using lattice QCD simulation a range of  values 0.1-0.4 \cite{Nakamura:2004sy}.
More recently for the $SU(3)$ gauge theory an estimate of $\eta/s\geq 0.134(33)$ is predicted at $T = 1.165 T_c$ \cite{0704.1801}. Therefore, a choice of $\eta/s\simeq0.1$ as representative is sensible. The proper time value has chosen to have this value in order to be larger than formation time and close to the lower bound of the thermalization time so to estimate the anisotropy effects  before the thermalization.

The formation time at the LHC is estimated lower than the RHIC, at least to $0.1 fm$ \cite{qgpbook}. Then by choosing a representative value $\t\simeq 0.5fm$ and $T=450 MeV$, relevant for LHC conditions we get $T \t=1.14$ and
\be\label{xilep}
\xi_L\simeq0.87~.
\ee
Therefore in order to compare between our anisotropic theory and the QCD, the last thing we need to specify is the relation between their proper times. One first approximation would be to consider proper times approximately equal. An other option is to fix the total entropy of motion since in our model the anisotropy parameter is very small. This assumption seems to be logical and in agreement with the discussion for heavy quarks in \cite{qgpbook}.
In that case the formation times between two theories with different degrees of freedom follow the relation
\be\label{tnormalized}
\t_{0\cN=4~sYM}=\left(\frac{T_{QCD}}{T_{\cN=4~sYM}}\right)^3
\left(\frac{d.o.f._{QCD}}{d.o.f._{\cN=4~sYM}}\right)\t_{0QCD}~.
\ee
Then we get for the formation times
\be
\t_{0\cN=4~sYM}\simeq 2.7^{-1/4} \t_{0QCD}
\ee
and for the $T\t$
\be
(T\t_0)_{\cN=4~sYM}=2.7^{-1/2} (T\t_0)_{QCD}~.
\ee
We see that both the formation time and the $T\t$ quantity is lower in $\N=4$ (anisotropic) sYM.
This implies for $\xi$ close to the initial conditions
\be
\xi_{\cN=4~sYM}\simeq\sqrt{2.7}~\xi_{QCD}~,
\ee
while the numerical values for the LHC and RHIC initial conditions we chose are translated to $\xi_{R~\cN=4~sYM} \simeq 4.60$  and  $\xi_{L~\cN=4~sYM}\simeq 3.78$.

We can assume that the proper time relation \eq{tnormalized}  between the two theories carry on beyond the formation time, for any proper early time we would like to compare:  $\t_{1 \cN=4~sYM}= 2.7^{-1/4}\t_{1 QCD}$  for our comparison purposes. For example this can be though as measuring the time with respect to the formation time in the two theories. In that case $\xi_{R~ aSYM}\simeq 2.14$ for the representative RHIC value \eq{xirep} and $\xi_{L~aSYM}\simeq 1.43$ for the representative LHC value \eq{xilep}.
Therefore from \eq{axi2} the resulting values for the supergravity anisotropy parameter are $(a/T)_R\simeq 5.81$ and $(a/T)_L\simeq 4.75$. For these values our current background can not be used for predictions since we are in $T\gg a$ approximation.

Notice if we would fix the entropy density instead of the energy density the relation between the formation times by making the above assumption of the constant total entropy would obviously be $\t_{0\cN=4~sYM}\simeq \t_{0QCD}$. This would lead to $(T\t_0)_{\cN=4~sYM}=2.7^{-1/3} (T\t_0)_{QCD}$ and consequently to $\xi_{\cN=4~sYM}\simeq2.7^{1/3}~\xi_{QCD}$. In this case to the values of $\xi$ parameters are very close to the ones found above, so qualitatively for our purposes there is no change.

Using the direct scheme where the temperatures of the two theories are identified, the isotropization time would be equal in the two  theories since in $\cN=4$ sYM it was found that $\t_{iso}\simeq 0.7/T$ \cite{yaffetim1}. Then the values of $\xi$ near the isotropization would coincide in the two theories and take the values \eq{xirep} and \eq{xilep} leading to $4.53$ and $3.7$ respectively. The $a/T$ values turn out to be greater than one and close to the values predicted above. So for our purposes the interesting outcome here is that by fixing different quantities in the two theories the proper times and more importantly the $T\t$ quantities do not differ significantly, and all of them lead to relatively high  $a/T$ values if we trust the equation \eq{xieta}.

To make the comparison using our model we should have had at most say $a/T\simeq 0.3$ which gives $\xi_{aSYM}\simeq 0.005$ which correspond to extremely high temperature according to equation \eq{xieta}. Therefore, we need to go to higher values of $a/T$ than our approximation allows. However, for higher values of  $a/T$ without considering the expansion, the pressure inequality $P_{x_3}<P_{x_1 x_2}$ get reversed and when this happens the supergravity solution does not describe anymore an expanding plasma with the desirable properties.
Notice that the values for the parameters we have used here are approximate, however any sensible values for the proper time (and how it is modified in the comparison scheme used) and the viscosity over entropy ratio that could be chosen would lead to the similar outcome for the range of values of large $a/T$ and anisotropy $\xi$.

The results we have obtained in our paper show how the particular observables are modified in presence of anisotropy in the strong coupling limit, and the qualitative pattern that they follow is  clear.
However for completeness we consider $a/T\simeq 0.3$ and plug its value to our results. Then the jet quenching modified as
\be
\frac{\hat{q}_{\parallel(\perp)}}{\hat{q}_0}\simeq 1.036~,\quad
\frac{\hat{q}_{ \perp(\parallel)}}{\hat{q}_0}\simeq 1.0009~,\quad
\frac{\hat{q}_{\perp(\perp)}}{\hat{q}_0}\simeq 1.0002~.
\ee
While the drag force, for velocity $v=0.95$ is
\be
\frac{F_\parallel}{F_\perp}\simeq 1.0035~,\quad
\frac{F_\parallel}{F_{iso}}\simeq 1.0036~,\quad
\frac{F_\perp}{F_{iso}}\simeq 1.0001~.
\ee

\section{Conclusions and List of Results}

In this paper we have studied the jet quenching, the drag force and the static potential, in the strong coupling dual anisotropic QGP. We have seen how certain quantities of the finite temperature isotropic $\cN=4$ sYM theory are modified in presence of anisotropy. In the QGP plasma, similar anisotropies can be created due to the expansion along the longitudinal direction. We have found several interesting results where some of them are listed briefly below:

$\bullet$ \textbf{Jet quenching:} In presence of anisotropy the jet quenching is enhanced compared to the isotropic case and its value depends strongly on the direction of the moving quark and the direction which the momentum broadening occurs. The jet quenching is strongly enhanced for a quark moving along the anisotropic direction and momentum broadening happens along the transverse plane. It reduces for a quark moving along the transverse direction and the momentum broadening considered along the anisotropic one. Finally, it reduces further when the quark moves in the transverse plane and the broadening happens on the same plane.
We can write our findings in a more compact form as
\be
\hat{q}_{\parallel(\perp)}>\hat{q}_{\perp(\parallel)}>\hat{q}_{\perp(\perp)}>\hat{q}_{iso}~. \ee
Additionally the dependence of $\hat{q}_{\parallel(\perp)}$ on the $a/T$ is linear in a very good approximation at small $a/T$ with
\be
\frac{\hat{q}_{\parallel(\perp)}}{\hat{q}_{0}}\simeq 1+ 0.122 \frac{a}{T}~,
\ee
but since normally is expected that the lower order term will be of second order the dependence becomes
\be
\frac{\hat{q}_{\parallel(\perp)}}{\hat{q}_{0}}\simeq 1+ 0.475 \left(\frac{a}{T}\right)^2~.
\ee

$\bullet$ \textbf{Drag Force:} In presence of anisotropy the longitudinal drag force $F_{drag,\parallel}$ is always enhanced. The transverse force $F_{drag,\perp}$ is enhanced above a critical velocity value $v\simeq 0.909$, while below this value is reduced. In both cases the anisotropic terms contributing to the anisotropy do not depend on the temperature \eq{fpar}, \eq{fper}. Comparing the forces between longitudinal and transverse directions we get
\be
\frac{F_{drag,\parallel}}{F_{drag,\perp}}=1+a^2 \frac{ \left(2-v^2\right) \text{Log}\left[1+\sqrt{1-v^2}\right]}{8 \pi ^2 T^2 \left(1-v^2\right)}~.
\ee
which indicates that $F_{drag,\parallel}>F_{drag,\perp}$~.

$\bullet$ \textbf{Diffusion time:} The diffusion time behavior is in direct analogy with all the findings of the drag force inverted. In the transverse direction $\t_\perp<\t_{iso}$ for $v\gtrsim 0.909$  while this relation is inverted for velocities lower than the critical value. The diffusion time along the longitudinal direction is always lower that the time in the isotropic medium. The relation between the times in the longitudinal and transverse directions are $\t_{\parallel}<\t_{\perp}$. The analytic relations for $\t_{\parallel}$, $\t_{\perp}$ normalized with the isotropic diffusion time are given by \eq{tdpar}, \eq{tdper} and their fraction by \eq{tdpp}.

$\bullet$ \textbf{Static Potential and Force:} We find that the static potential in presence of anisotropy becomes weaker in absolute value. The relation of the static potential of a pair aligned longitudinal to the anisotropy compared to the static potential of a pair aligned transverse direction is \be
V_\parallel<V_\perp<V_{iso}
\ee
when the comparison is done in terms of $L T$  keeping the anisotropy parameter and temperature fixed. Moreover the critical length of the string is decreased in presence of anisotropy as
\be
L_{c\parallel}<L_{c\perp}<L_{c~iso}~.
\ee
In order to get rid of the non-physical constant in the potential we consider the static force, and we find that indeed the force is screened in presence of anisotropy with the order
\be
F_{Q\bar{Q},\perp}<F_{Q\bar{Q},\parallel}<F_{iso}
\ee
and that increase of anisotropy leads to further decrease of the static force.

The modifications of most of our results due to anisotropy are stronger along the anisotropic direction. This happens because we have found that the geometry is modified more in the anisotropic direction than in transverse one, phenomenon that gets even stronger as the anisotropy parameter is increased. The geometry modifications reflect to the observable results. Moreover, for larger anisotropies where also the pressure anisotropy \eq{pressaniso} is inverted the  behavior of the observables we have found here is expected in some cases to be different. It should also be noted that the comparison of observables between anisotropic and isotropic theories have been done mostly by identifying the temperatures in these theories.

It is worth noticing that for large velocities the drag force and jet quenching parameter have the same qualitative behavior. This qualitative agreement has been observed in other cases too. In \cite{0606134}, the fraction of drag forces and jet quenching in cascading plasma or charged plasma over the $\N=4$ sYM have been found to follow similar patterns. for completeness we point out that exact numerical agreement for the fractions $\hat{q}_{\parallel(\perp)}/\hat{q}_{0}$ and $F_{drag,\parallel}/F_{0}$ can be found
for velocities $v\simeq 0.9996$.
Moreover it has been found that the presence of R-charges enhances the jet quenching \cite{sfetsos}. Therefore,  the inclusion of anisotropy and R-charges is expected to lead to further enhancement.

The particular enhancement in jet quenching due to the anisotropy and the longitudinal, transverse relations $\hat{q}_{\perp(\parallel)}>\hat{q}_{\perp(\perp)}$ we have obtained are in partial agreement with several models for weak coupling plasmas, eg. \cite{jetqa2,jetsu2,jetqa1} as well as with the STAR findings \cite{star}.
The static potential findings are different than the ones obtained in weak coupling limit, but in a different setup in \cite{qqaniso} where the potential was calculated from the Fourier transform of the static gluon propagator in a hard loop approximation. Moreover in  \cite{0901.1998} for the weakly anisotropic plasmas it was also found that the quarkonium binding is stronger for non-vanishing viscosity and expansion rate, result that is in agreement with \cite{qqaniso}. In strong coupling limit using our model where no dynamical flavor degrees of freedom,  we find weaker potential due to anisotropy. Although the inclusion of flavors beyond probe (unquenched) approximation in an isotropic theory is expected to lead to further screening  \cite{giataganasSS,qqscreened} it is not clear a priori what happens in case of presence of anisotropy. We have commented on the reasons in the section 3.1.

Moreover, using some sensible comparison schemes we have tried to make a more `quantitative' predictions and comparison with the experimental results. However,  the limit we have considered $T\gg a$ and the background is analytically known and has the desirable pressure anisotropy along the different directions, corresponds to low values of anisotropic parameter $\xi$, that seem not to be the ones that the observed QGP has, if we trust the equation \eq{xieta}.

There are several other studies that can be performed using the AdS/CFT for anisotropic QGP. The quarkonium physics, where the introduction of flavor branes are needed in order to insert to the background information for the dynamical degrees of freedom and the mesons. The electromagnetic observables are also interesting to be studied and is believed that they contain information about the initial stages of the anisotropic QGP. An effort to this directions has been done in \cite{electromagnetic} using the background \cite{janikaniso}. The photon and dilepton thermal production by the plasma has been studied in the weak coupled anisotropic plasma in \cite{photon}. Very interesting is also the derivation  of supergravity anisotropic solutions which depend on the time and describe the isotropization of the plasma.

It would be also interesting to see if and how the generic conditions for the cancelation of the UV divergences in the Wilson loops with the use of the Legendre transform for isotropic backgrounds, derived in \cite{giataganasWL} are modified in case of anisotropic backgrounds. In the special case of the orthogonal static potential Wilson loop, it seems that the result of the UV divergence with the use of the Legendre transform, gives the same formula as the isotropic case. However, for more complicated Wilson loops the analysis is more involved.

\textbf{Notice:}

The results of this paper were reported in a talk, at 31 January of 2012 in the workshop `Exploring QCD frontiers: from RHIC and LHC to EIC', Stellenbosch 2012. 
While this paper was in final stage of typing for submission the author has received the reference \cite{mateosdr} which has small partial overlap with our drag force results, and in particular there is an agreement for small $a/T$ limit. Moreover, some time after the appearance of our paper the same authors have studied the jet quenching for generic quark motion and for the whole range of $a/T$ \cite{Mateosjet}. In the small $a/T$ limit their results again agree with ours. For larger anisotropies $a/T$, where however the pressure inequality \eq{pressaniso} is inverted, the jet quenching behavior changes in some directions.

\textbf{Acknowledgements:}

The author would like to thank R. de Mello Koch, K. Goldstein, F. Knechtli, K. Siampos, H. Soltanpanahi and the participants of the workshop `Exploring QCD frontiers: from RHIC and LHC to EIC', Stellenbosch 2012, for very useful discussions. The research of the author is supported by a Claude Leon postdoctoral fellowship. The author also participates in PEVE 2010 NTUA program.

\startappendix

\Appendix{Q\={Q} Strings in generic weakly coupled backgrounds}

In this section we present the world-sheet calculation of a string in static gauge in weakly coupled backgrounds and their energy which corresponds to the static potential.
The string in the anisotropic background \eq{metric}, is a special case of the strings we examine here. The following equations have been derived in various forms in several in other papers too \cite{stringtension,giataganasSS}.
By writing the metric of the space as
\be\label{metricqq}
ds^2=G_{00}d\t^2+G_{ii}dx_i^2+G_{uu}du^2~,
\ee
we choose  the static gauge for the string
\be
x_0=\t\qquad\mbox{and}\qquad x_p=\sigma,
\ee
which is extended in the radial direction, so $u=u(\s)$. The $x_p$ coordinate represents the direction along which the pair is aligned and can be chosen to be
\bea\nn
&&x_p=x_{1,2}=:x_{\perp} \quad\mbox{or}\\
&&x_p=x_3=:x_{\parallel}~.\label{xppp}
\eea
Then the induced metric $g_{\alpha \beta}=G_{MN}\partial_\alpha X^M \partial_\beta X^N$ for our world-sheet read:
\bea
g_{00}=G_{00},\qquad g_{11}=G_{pp}+G_{uu}u'^2.
\eea
Supposing that we are working in Lorentzian signature the Nambu-Goto action is\footnote{In the case of Euclidean signature, the formulas change with a minus sign wherever the $G_{00}$ element is.}
\bea
S=\frac{1}{2\pi\a'}\int d\s d\t \sqrt{- G_{00}(G_{pp}+G_{uu}u'^2)}=:\frac{1}{2\pi\a'}\int d\s d\t \sqrt{D}~.
\eea
The Hamiltonian then is equal to
\be
H=\frac{ G_{00} G_{pp}}{\sqrt{D}}
\ee
and is a constant of motion. Setting it equal to $-c$ we can solve for $u'$ and get the turning point equation
\be\label{uprime0}
u'=\pm\sqrt{-\frac{(G_{00} G_{pp}+ c^2)G_{pp}}{ c^2 G_{uu}}}~,
\ee
which is solved for
\be\label{tpsol}
G_{uu}^{-1}=0~,\qquad\mbox{or}\qquad G_{pp}=0~,\qquad\mbox{or}\qquad G_{00}G_{pp}=- c^2~.
\ee
The above equations, normally the last one, specify how deep the world-sheet goes into the bulk, and we call this value of the turning point $u_{0}$.

The length\footnote{The limits of the length integral depend on where we choose as the starting point measuring $L$. When the string in the boundary extends from $-L/2$ to $L/2$, then the corresponding solution of the $u'$ is positive for $(0,L/2)$ since the turning point corresponds to $L=0$. When the string extends from $(0,L)$ the $u'$ in $(0,L/2)$ is negative. In any case the final result in the definite integral is \eq{staticL}.} of the two endpoints of the string on the brane is given by
\be\label{staticL}
L=2\int_{\infty}^{u_{0}}\frac{du}{u'}=2\int_{u_{0}}^{\infty}  du \sqrt{\frac{- G_{uu} c^2}{(G_{00}G_{pp}+ c^2)G_{pp}}}~.
\ee
Moreover, the energy of the string using as renormalization method the mass subtraction of the two free quarks is
\bea\nn
2\pi\a' E&=&2\left(\int_{u_{0}}^{\infty} d\s \cL -\int_{u_{k}}^{\infty} du \sqrt{ G_{00}G_{uu}}\right)\nonumber\\\nn
&=& c L+2\left[  \int_{u_{0}}^{\infty} du \sqrt{- G_{uu}G_{00}}\left(\sqrt{1+\frac{c^2}{G_{pp}G_{00}}}-1\right)- \int_{u_{k}}^{u_0} du \sqrt{-G_{00}G_{uu}}\right]~,\\\label{staticE}
\eea
where $u_k$ is the possible horizon of the metric.
Notice that we already used that fact that the world-sheet is symmetric with respect to turning point $u_0$ and hence the RHS of the above equation are already multiplied by two.

Using the equations \eq{staticL} and \eq{staticE} the static potential can be found in terms of the distance of the Q\={Q} pair for any gravity dual background at least numerically. The derivation of the analytical expressions depend on whether the integrals can be done analytically and if the inversion of $u_0(L)$ is possible analytically.

\Appendix{Drag Force on trailing string in general weakly coupled backgrounds}\label{app:drag}

It is possible to calculate the drag force analytically in a generic background, for example as in \cite{kiritsis}, and as we have done for the static potential. We consider the background with the generic metric \eq{metricqq} and  the following ansatz for the trailing string with radial gauge choice and motion along the $x_p$ direction:
\be
x_0=\t, \qquad u=\s, \qquad x_p= v \t +f(u)~,
\ee
where in the other directions the world-sheet is localized. The $x_p$ is chosen to be $x_{\parallel}$  or $x_\perp$, like the \eq{xppp} for the static potential analysis. It denotes the direction along which the heavy quark moves with velocity $v$.  Moreover, the function $f(u)$ at the boundary should be equal to zero in order to resemble a constant quark motion.
The Nambu-Goto action in our case reads
\be
S=-\frac{1}{2\pi\a'} \cT \int du\sqrt{-\left(G_{00}+G_{pp}v^2\right) G_{uu}-G_{00} G_{pp}f'^2}=:-\frac{1}{2\pi\a'} \cT \int du\sqrt{D_1}~.
\ee
The action produces one non-trivial equation for $f$ and since it does not depend explicitly on $f$, it has a constant of motion the canonical momentum
\be\label{momp}
\Pi_u^1=\frac{1}{2\pi\a'}\frac{G_{00} G_{pp}f'}{\sqrt{D_1}}~.
\ee
The $f'$ function must be real and to elaborate on the reality condition we need to solve \eq{momp} for $f'$ to get
\be\label{dragxi}
f'=\frac{\sqrt{-\left(G_{00}+G_{pp}v^2\right) G_{uu}}}{\sqrt{-G_{00} G_{pp}\left(1+ G_{00} G_{pp}\left(2 \pi \a'\Pi_u^1 \right)^{-2}\right)}}~.
\ee
To keep $f'$ real, both the numerator and denominator must have the same sign at any value $u$. The numerator vanishes for $u=u_0$ satisfying
\be\label{dragu0}
G_{uu}(G_{00} +G_{pp} v^2)~\mid_{u=u_0}=0~,
\ee
where usually the solution comes from the expression within the brackets.
The momentum then can be expressed in terms of $u_0$ where the denominator simplified significantly
\be\label{dragcond}
\Pi_u^1=-\frac{\sqrt{-G_{00} G_{pp}}}{2 \pi \a'}\bigg|_{u=u_0}~.
\ee
Since we have chosen the physical solution which describes the momentum flowing along the string from the boundary to the horizon, the total drag force on the string  for motion along the $x_p$ direction is given by
\be\label{drag}
F_{drag,x_p}=\Pi_u^1=-\sqrt{\l}\frac{\sqrt{-G_{00} G_{pp}}}{(2 \pi)}\bigg|_{u=u_0}~.
\ee
Therefore in any generic background  the drag force can be calculated  by solving \eq{dragu0} to specify the point $u_0$ and then by plugging its value to the equation \eq{drag}. 

\end{document}